\documentclass[preprint]{aastex63}

\usepackage{lineno,hyperref}
\usepackage{amssymb}
\usepackage{graphicx}
\usepackage{array}
\usepackage{sidecap}
\usepackage[section]{placeins}

\hypersetup{linkcolor=red,citecolor=green,filecolor=cyan,urlcolor=blue}

\received{Aug. 5, 2019}
\revised{Oct. 21, 2019}
\accepted{Oct. 30, 2019}
\submitjournal{AJ}

\shorttitle{HST Outburst Imaging of SW1}
\shortauthors{Schambeau et al.}

\begin{document}

\title{Analysis of HST WFPC2 Observations of Centaur 29P/Schwassmann-Wachmann 1 while in Outburst to Place Constraints on the Nucleus' Rotation State}

\correspondingauthor{Charles Schambeau}
\email{charles.schambeau@ucf.edu}

\author[0000-0003-1800-8521]{Charles A. Schambeau}
\affiliation{Florida Space Institute, University of Central Florida \\
12354 Research Parkway, Partnership 1 \\
Orlando, FL 32826, USA}

\author[0000-0003-1156-9721]{Yanga R. Fern\'andez}
\affiliation{Department of Physics, University of Central Florida \\
Orlando, FL 32816, USA}
\affiliation{Florida Space Institute, University of Central Florida \\
12354 Research Parkway, Partnership 1 \\
Orlando, FL 32826, USA}

\author[0000-0001-8925-7010]{Nalin H. Samarasinha}
\affiliation{Planetary Science Institute \\
Tucson, AZ 85719, USA}

\author{Laura M. Woodney}
\affiliation{Department of Physics, California State University, San Bernardino\\
San Bernardino, CA 92407, USA}

\author{Arunav Kundu}
\affiliation{Eureka Scientific\\
Oakland, CA, USA}



\begin{abstract}

We present analysis of {\it Hubble Space Telescope} (HST) observations of Centaur 29P/Schwassmann-Wachmann 1 (SW1) while in outburst to characterize the outburst coma and place constraints on the nucleus' spin state. The observations consist of Wide Field and Planetary Camera 2 (WFPC2) images from Cycle 5, GO-5829 (\citet{1995hst..prop.5829F}) acquired on UT 1996 March 11.3 and 12.1, which serendipitously imaged the Centaur shortly after a major outburst. A multi-component coma was detected consisting of: an expanding outburst dust coma with complex morphology possessing an east-west asymmetry and north-south symmetry contained within 5$''$ ($\sim$19,000 km) of the nucleus, the residual dust shell of an earlier UT 1996 February outburst, and a nearly circular underlying quiescent activity level coma detectable to $\sim$70$''$ ($\sim$267,000 km) away from the nucleus. Photometry of the calibrated WFPC2 images resulted in a measured 5$''$ radius aperture equivalent R-band magnitude of 12.86 $\pm$ 0.02 and an estimated (2.79 $\pm$ 0.05)$\times 10^8$ kg for the lower limit of dust material emitted during the outburst. No appreciable evolution of morphologic features, indicating signatures of nucleus rotation, were detected between the two imaging epochs. The observations were modeled using a 3-D Monte Carlo coma model (\citet{2000ApJ...529L.107S}) to place constraints on the nucleus' rotation state. Modeling indicated the morphology is representative of a non-isotropic ejection of dust emitted during a single outburst event with a duration on the order of hours from a single source region corresponding to $\sim$1\% of the surface area. A spin period with lower limit on the order of days is suggested to reproduce the coma morphology seen in the observations. 

\end{abstract}

\keywords{general, centaurs --- 
individual, 29P/Schwassmann-Wachmann 1 --- {\it Hubble Space Telescope} --- optical observations --- outbursts}


\section{Introduction}
Centaurs are small-icy bodies in orbits between Neptune and Jupiter, and represent intermediaries in the dynamical link between the outer solar system's Trans-Neputnian Objects (TNOs; more specifically the scattered disk objects) to the inner solar system's Jupiter-Family comets (JFCs) (\citet{1997Sci...276.1670D}; \citet{2004come.book..193D}). In situ observations of spacecraft-visited JFCs nuclei have revealed highly evolved bodies which have undergone complex thermal processing (\citet{sunshine_2016}) and the New Horizons Kuiper Belt Extended Mission's flyby of the TNO 2014 MU$_{69}$ (\citet{2018SSRv..214...77S}) has given us an example of the ``pristine'' materials we can expect to populate the cold storage environment of the outer solar system. Although 2014 MU$_{69}$ is in the classical cold Kuiper Belt, it represents an example of the types of cryogenically stored materials potentially contained in the scattered disk of progenitor small-icy bodies that supply the JFCs population. To fully understand the evolutionary link between the preserved outer Solar System materials and the more easily observationally accessible but also more thermally processed JFCs, we must understand the material evolution ongoing in the Centaur region. Observations of active Centaurs have shown that the active lifetimes of JFCs often start in this region (\citet{2007MNRAS.381..713M}; \citet{2009AJ....137.4296J}), indicating ongoing material processing in the Centaur region. The Centaur 29P/Schwassmann-Wachmann 1 (SW1) is an observational target at the cusp of the Centaur-to-Jupiter-Family transition region (\cite{2019ApJ...883L..25S}) and presents a rare opportunity to investigate the activity drivers and ongoing material processing occurring in the Centaur region.

SW1 is an enigmatic object due to the combination of its unique orbital properties and its activity behaviors: (1) Its orbit is nearly circular just beyond the orbit of Jupiter (eccentricity = 0.043, semi-major axis {\it a} = 6.026 AU, and inclination {\it i} = 9.368$^{\circ}$ (IAU Minor Planet Center, Epoch 2019-03-18.0 TT) and at the time of writing of this manuscript, has the third lowest eccentricity of known comets or active Centaurs. (2) It is continuously active in a region too cold for rigorous water-ice-sublimation driven activity. This continuous activity, often called its ``quiescent" activity, is believed to be driven by sublimation of the supervolatile CO (\citet{2001Icar..150..140F}; \citet{2002Icar..157..309G}; \citet{2013ApJ...766..100P}) and/or from gases released during the crystallization of amorphous water ice (\citet{2009AJ....137.4296J}; \citet{2017PASP..129c1001W}). (3) Additionally, it frequently, and consistently, undergoes major outbursts events in activity with estimated lower-limits for the total mass of material ejected on the order of $10^8 - 10^9$ kg per outburst event (\citet{2010MNRAS.409.1682T}; \citet{2013Icar..225..111K}; \citet{2013AJ....145..122H}; \citet{2016Icar..272..327M}; \citet{2017Icar..284..359S}). These outburst events are detected as corresponding increases in surface brightness of the coma's dust continuum features and range over several orders of magnitude, lasting from several days to a few weeks. With SW1's nearly circular orbit providing an environment with nearly constant insolation, the question arises as to what are the ongoing processes causing its varying stages of activity? With a seemingly stable thermal environment, why would the nucleus experience frequent excursions from a steady state of activity? Does SW1's enigmatic activity represent the behaviors of pristine outer solar system material being thermally activated during its initial passages into the inner solar system? Measurements of SW1's CO-production rate relative to its nucleus' size and dust-activity behaviors indicate it has more similarities with long-period comets (e.g. Comet Hale-Bopp) than to most JFCs (\citet{2015ApJ...814...85B}; \cite{2017AJ....153..230W}; \citet{2017PASP..129c1001W}), supporting the notion that SW1 possesses more pristine materials. Our lack of fundamental understanding of SW1's behavior and the high relative level of activity present at a relatively large heliocentric distance suggests a recent arrival from the outer solar system, emphasizing SW1 as a high-priority target for continued observations and analysis to gain a better understand of the Centaur-to-Jupiter-Family transition region. 

One method of investigating the ongoing material processing of active bodies and also to probe the compositions and structures of the sub-surface layers of a nucleus' interior is through thermophysical modeling of the body's nucleus (\citet{2004come.book..359P}; \citet{2004come.book..317M}). These modeling efforts are constrained by dust- and gas-production rates measured from observations at different epochs along the object's orbit. Knowledge of the nucleus' spin state, locations of surface regions of activity, and the nucleus' shape allow an accurate representation of the insolation distribution received by the nucleus throughout its orbit and are vitally important for successful thermophysical modeling efforts. For SW1 there have been many attempts to determine its nucleus' spin state, but no consistent set of constraints have been determined. Spin period estimates on the order of several hours, similar to measurements for spin states for an ensemble of JFCs (\citet{2017MNRAS.471.2974K}), have been derived from analysis of photometric light curves measured using small photometric apertures centered on the nucleus' position (\citet{1993AJ....106.1222M}) and also through analysis of coma morphology detected in {\it Spitzer} IRAC 4.5 $\mu$m images (\citet{2013Icar..226..777R}). Alternatively, spin period estimates on the order of $\sim$50 days have been proposed from analysis of trends in SW1's possible outburst periodicity (\citet{2010MNRAS.409.1682T}; \citet{2016Icar..272..327M}; \cite{2016Icar..272..387M}). No definitive spin state constraints exist for this enigmatic Centaur, limiting our ability to fully investigate the nature of its activity patterns through thermophysical modeling efforts.

Our group is investigating SW1's nucleus spin state through analysis and modeling of outburst dust-comae morphology detected in broadband dust continuum imaging. We are using a 3-D Monte Carlo coma model to generate synthetic observations and a comparative procedure between observations and synthetic observations to place constraints on possible spin states that can replicate the observations. Our first such analysis is presented in \citet{2017Icar..284..359S} where we placed constraints on the ratio of spin period to outburst duration (i.e. the length of time over which increased dust lofting from the nucleus occurred due to the outburst event). From this analysis it was determined that a lower limit for the nucleus' rotation period on the order of several days was required to replicate the minimal rotation of the nucleus that occurred during the 2008 outburst event, which was constrained by the lack of detectable evolution of features present in the outburst dust coma's morphology.

In this follow-on article we present an analysis of {\it Hubble Space Telescope} (HST) observations of SW1 shortly after a major outburst event in 1996, where we further constrain the nucleus' spin state. Section 2 gives details of the 1996 HST observations including a review of the results derived from the analysis undertaken by \citet{1996DPS....28.0808F}. In Section 3 we describe our image analysis, which predominantly included the application of a suite of image enhancement routines developed for application with cometary comae images (\citet{2014Icar..239..168S}) and estimates on the amount of dust material emitted during the outburst event. Section 4 introduces the topic of Monte Carlo coma modeling and application of this analysis technique to the 1996 HST observations. Finally, in Section 5 we summarize the results of the coma modeling analysis and implications the derived nucleus spin state have on understanding the activity behaviors of SW1. 

\section{Observations} \label{sec:observations}
The observations analyzed by our group are Wide Field and Planetary Camera 2 (WFPC2) images from HST Cycle 5, Program GO-5829 ``The Activity of Periodic Comet Schwassmann-Wachmann 1" (\citet{1995hst..prop.5829F}) and were retrieved for our analysis from the Mikulski Archive for Space Telescopes (MAST). The primary science objectives of the GO-5829 program focused on spectroscopic observations using the Faint Object Spectrograph (FOS) and Goddard High-Resolution Spectrograph (GHRS) to investigate the volatile species driving SW1's activity. Initial analysis of the observations from their group resulted in estimated upper limits for the water-production and CO-production rates (\citet{1996DPS....28.0808F}): [$Q$]$_{\textrm{H}_2\textrm{O}} < $ 3x10$^{28}$ molecules/second and [$Q$]$_{\textrm{CO}} < $ 9x10$^{28}$ molecules/second. The WFPC2 observations were acquired before acquisition of the first FOS spectrum and after the last FOS spectrum. Details of the WFPC2 observations are provided in Table \ref{tab:geometry}. Two epochs of four WFPC2 exposures were acquired during two HST orbits; the individual orbits were separated by $\sim$19.3 hours. Hereafter images are referred to by their observational epoch as follows: ``Exp-1" identifies images captured during the Mar. 11 orbit and ``Exp-2" identifies images captured during the Mar. 12 orbit. 

\begin{deluxetable*}{ccccccccc}
\tablenum{1}
\label{tab:geometry}
\tablecaption{SW1 Observations Summary for UT March, 1996}
\tablewidth{0pt}
\tablehead{
\colhead{Date} & \colhead{Observation ID} & \colhead{Observation Start} & \colhead{Phase Angle} & \colhead{[R$_H$, $\Delta$]$^{\textrm{a}}$} & \colhead{Exp. Time$^{\textrm{b}}$} & \colhead{Filter$^{\textrm{c}}$} & \colhead{$\rho_{\textrm{Sat}}$$^{\textrm{d}}$} \\
\colhead{} & \colhead{} & \colhead{(UT)} & \colhead{(deg)} & \colhead{(au)} & \colhead{(s)} & \colhead{} & \colhead{($''$)}
}
\startdata
		11	&	u3490101t	&	07:38:15			&	2.34			&	[6.262, 5.296]					&		120				&	F702W			&	0.25								\\
		11	&	u3490102t	&	07:42:16			&	2.34			&	[6.262, 5.296]					&		400				&	F702W			&	0.50								\\
		11	&	u3490103t	&	07:51:16			&	2.34			&	[6.262, 5.296]					&		600				&	F702W			&	0.60								\\			
		11	&	u3490104t	&	08:08:16			&	2.34			&	[6.262, 5.296]					&		600				&	F702W			&	0.60								\\
		12	&	u3490401p	&	02:56:15			&	2.47			&	[6.262, 5.300]					&		120				&	F702W			&	--								\\		
		12	&	u3490402p	&	03:00:16			&	2.47			&	[6.262, 5.300]					&		400				&	F702W			&	0.08								\\		
		12	&	u3490403p	&	03:09:16			&	2.47			&	[6.262, 5.300]					&		600				&	F702W			&	0.14								\\		
		12	&	u3490404p	&	03:26:16			&	2.47			&	[6.262, 5.300]					&		600				&	F702W			&	0.14								\\
\enddata
\tablecomments{\\
$^{\textrm{a}}$ Heliocentric and HST distance during observation (Horizons, JPL). \\
$^{\textrm{b}}$ Exposure time for each image frame.\\
$^{\textrm{c}}$ WFPC2 filter selected for the observation.\\
$^{\textrm{d}}$ Projected radii of regions containing pixel saturation centered on Horizons ephemeris position of the nucleus.}
\end{deluxetable*}

\begin{figure}
\gridline{
		\fig{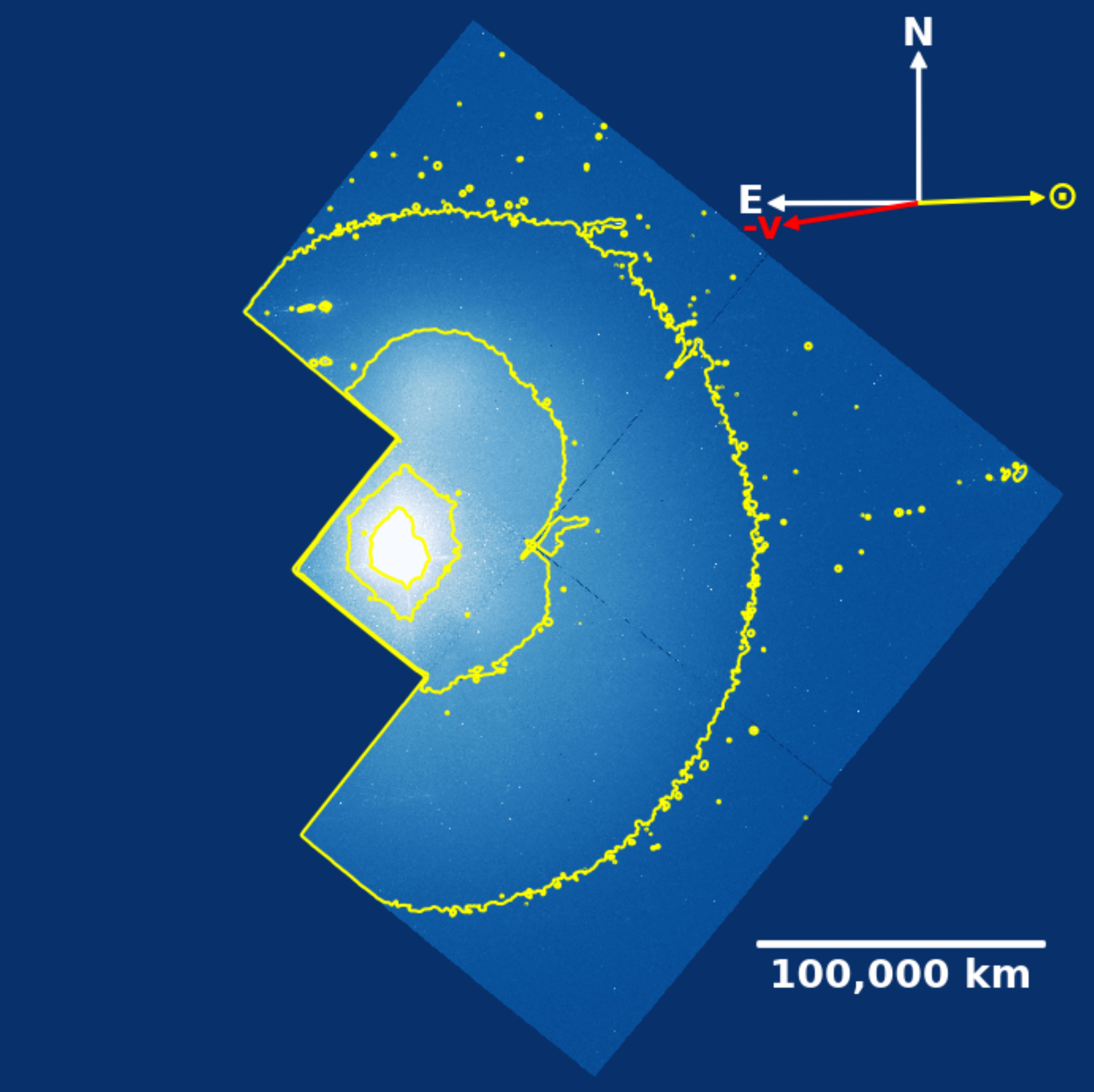}{0.5\textwidth}{(a)}
          	\fig{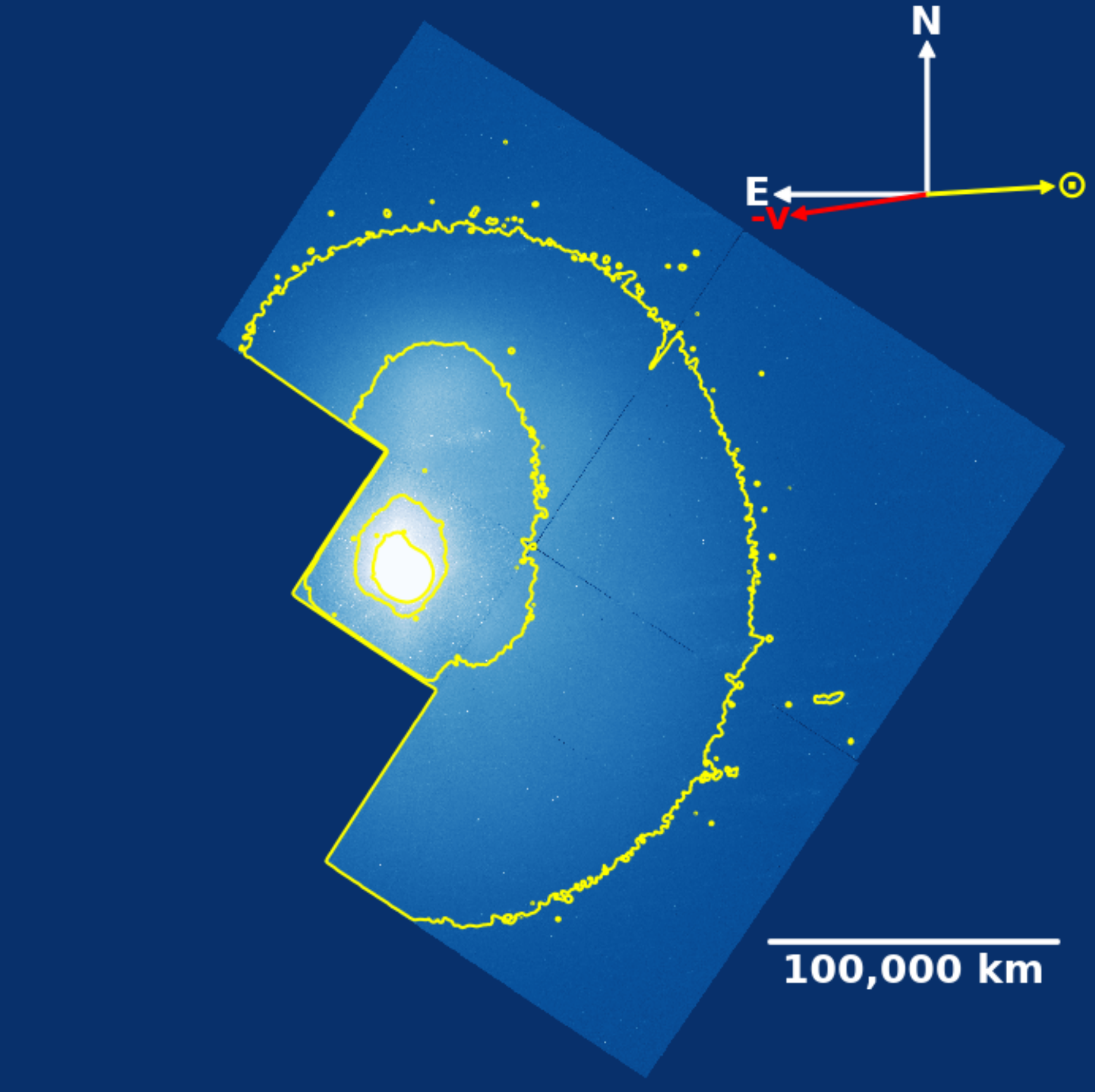}{0.5\textwidth}{(b)}
          }
\caption{WFPC2 observations of SW1 from 1996. (Left panel) Resultant image after median-combining the four mosaicked images from epoch Exp-1. (Right panel) same as left but for epoch Exp-2, acquired 19.3 hours later. Exp-1 is saturated in the region of the image containing the nucleus. In both observations equatorial north and east are indicated. The position angles of the Sun and negative Centaur velocity vector are indicated by the yellow and red arrows. A scale bar on the bottom right of each image shows a projected distance of 100,000 km at the location of SW1. The yellow contours show the outline of the quiescent coma's contribution extending onto the wide field cameras. The small localized contours in the images are the result of artifacts in the image and are not surface brightness variations inherent to the coma.  \label{fig:WFPC2}}
\end{figure}

WFPC2 was in operation from 1993 December until 2009 May. Details of WFPC2 and analysis of its data products can be found in the Instrument Handbook (\citet{2008hst...macmaster}) and Data Handbook (\citet{2010hst...gonzaga}). The camera imaged an $\sim$160$''$ x $160''$contiguous region onto four individual cameras by splitting the incoming beam by reflection off a four-faceted pyramid mirror. The four cameras are: the Planetary Camera (PC) and three Wide Field Cameras (WFC; individual cameras are referred to as WF2, WF3, and WF4). A single exposure using the WFPC2 comprised simultaneously acquiring images from each of the four cameras. The PC had a pixel scale 0$''$.046/pixel and a 36$''.8$ x 36$''$.8 field-of-view (FOV), while the three WF cameras had a pixel scale of 0$''.1$/pixel and an 80$''$x 80$''$ FOV. Figure \ref{fig:WFPC2} shows examples of the individual WFPC2 images acquired during a single exposure after being mosaicked together and the PC image binned to an effective 0$''$.1/pixel scale. For our analysis, we used individual calibrated science data products generated from the OPUS software system version 2008\_4 and CALWP2 code version 2.5.3 (calibration performed on September 4, 2008; images with \_c0f.fits extension) and the quick-look drizzled mosaic images generated by PyDrizzle version 6.3.0 (calibration on September 3, 2008; images with \_drz.fits extension). The single epoch WFPC2 images (Exp-1 or Exp-2) shown in Figure \ref{fig:WFPC2} are the result of median combining the four individual level 2 mosaicked and calibrated observations retrieved from MAST for each of the two observing epochs.     

The GO-5829 observations serendipitously captured the Centaur shortly after a major outburst. During the observations, SW1's nucleus was positioned near the center of the PC camera. The unexpected nature of the outburst observations unfortunately resulted in all PC images, except the 120 second Exp-2 PC image (Observations ID: u3490401p), to be saturated in pixels close to and centered on the nucleus. Table \ref{tab:geometry} includes the extent of pixel saturation for each image. While the saturated images have a reduced image quality when compared to images acquired with proper exposures, capturing observations of SW1 using HST only hours after a major outburst have resulted in the highest resolution images of SW1's outburst coma to date. These high-resolution images, with increased signal-to-noise coma surface brightness detections due to the increased effective dust scattering cross-section, are ideally suited for coma morphology analysis and Monte Carlo coma modeling for nucleus spin state constraints. The lag time between initial notification for a target-of-opportunity (ToO) program and the acquisition of ToO observations would have required a longer time interval between outburst onset and first image acquisition. ToO HST observations of this event would have been hard pressed to logistically capture an outburst so shortly after the currently unpredictable and short-lived event.

\section{Image Analysis}

\subsection{Outburst Coma Morphology} \label{sec:morphology}
\begin{figure}[ht!]
\gridline{
		\fig{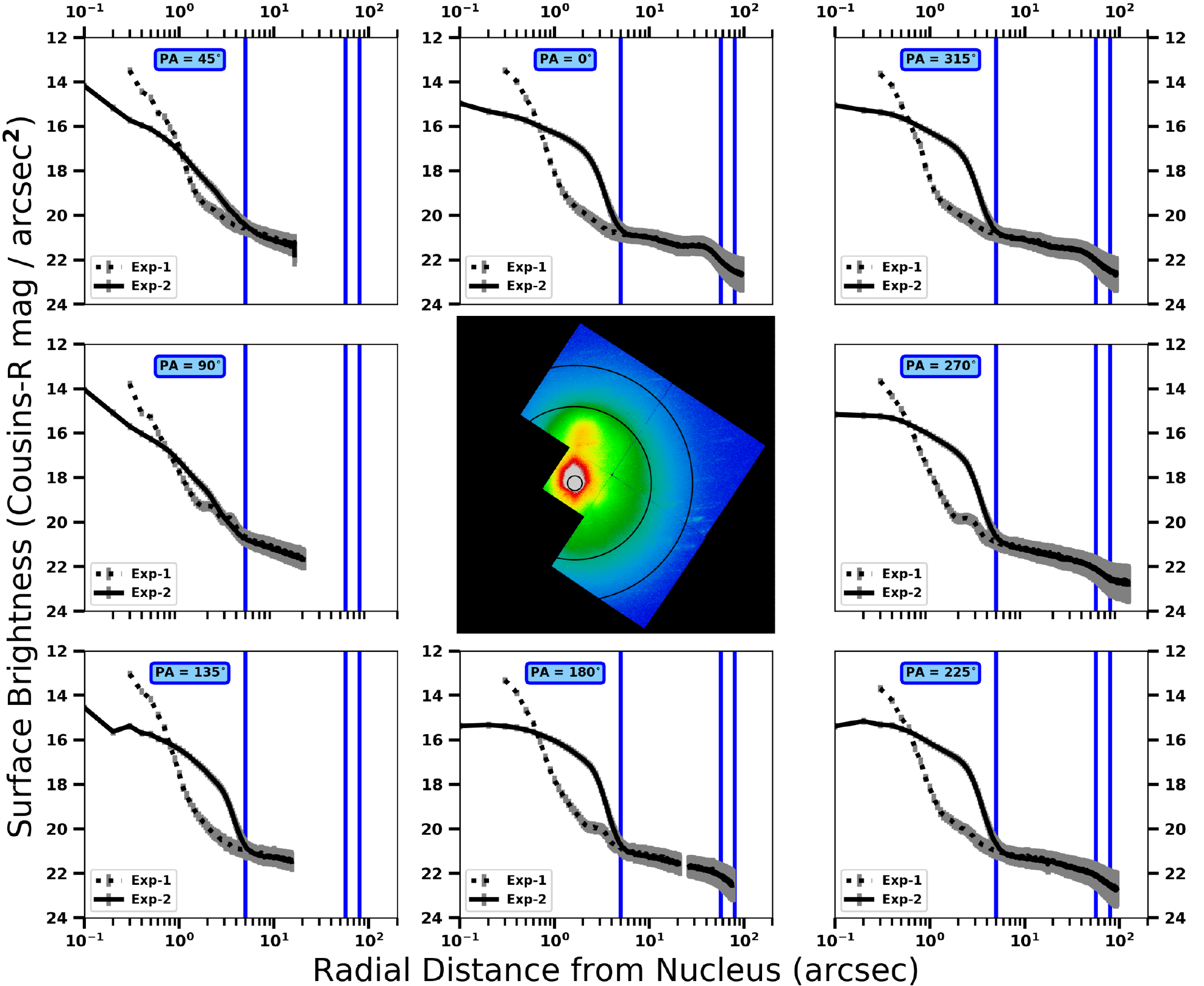}{1.0\textwidth}{}
          	}
\caption{Radial surface brightness profiles of the two WFPC2 images shown in Figure \ref{fig:WFPC2}. The image in the center panel is the Exp-2 WFPC2 observations from Figure \ref{fig:WFPC2} (right panel) displayed with a different colormap to highlight the morphology of the coma. Black circles overlain on the image represent the positions of the blue vertical lines in each plot for projected cometocentric distances $\rho$ = 5$''$.0, 57$''$, and 80$''$ where the comae slopes change appreciably. The position angles of the radial surface brightness profiles are indicated in each plot. The overlap between Exp-1 and Exp-2 profiles for each PA at radial distances larger than 5$''$ indicates that minimal evolution occurred for this portion of the coma between the two epochs of observations and also that the dust emitted during the recent outburst is contained within a 5$''$ radius aperture centered on the nucleus. The Exp-1 profiles interior to 0$''$.3 are excluded from the plots due to pixel saturation. Regions of localized enhanced surface brightness for Exp-1 profiles between 2-3$''$ for PAs 90$^{\circ}$, 180$^{\circ}$, and 270$^{\circ}$ are the result of diffraction spike artifacts due to the Exp-1's overexposure.  \label{fig:WFPC2_Profiles}}
\end{figure}

The WFPC2 images show a complex, multi-component dust coma composed of both steady-state quiescent activity and outburst dust comae. \citet{1987ApJ...317..992J} have shown the abundance of information that can be extracted from comae surface brightness profiles and we apply these techniques as a first-order approach to understanding the underlying activity of SW1's nucleus. Figure \ref{fig:WFPC2_Profiles} shows radial surface brightness profiles centered on the nucleus' position for both epochs of imaging for position angles (PA) at 45$^{\circ}$ spacings. The pixel saturation inherent in the Exp-1 images required estimating the location of SW1's nucleus in the image. This was accomplished by using a 100$\times$100 pixel region surrounding the region of Exp-1 pixel saturation. This cropped region was separated into two halves about the middle array column. Individual rows in each half of the image were fitted with a 1-D polynomial of degree 10 for regions of the outburst coma outside of the region of pixel saturation. A polynomial of degree 10 was chosen to allow sufficient degrees of freedom to replicate the complex surface brightness profile behavior of the outburst coma. This row-fitting procedure was performed on the western and eastern halves of the image separately. The fitting-returned polynomials were used to extrapolate the coma's surface brightness behavior into the region of saturation. A saturation-filled outburst coma model was created by joining the western and eastern halves of the corrected images. A similar procedure was implemented by splitting the image into northern and southern halves and fitting the column pixel values around the region of saturation. The final simulated Exp-1 inner outburst coma used was determined by taking the average of the extrapolated pixel values from the two simulated inner coma models. The location of the maximum pixel value in the 2-D fitted coma surface brightness distribution was used as the location of SW1's nucleus for Exp-1 radial surface brightness profiles. For Exp-2, the nucleus' position was chosen to be located at the brightest pixel's location. The radial profiles were produced by median combining pixel values from 10$^{\circ}$-wide pie-shaped wedges centered on the indicated PA. 

\begin{figure}[ht!]
\gridline{
		\fig{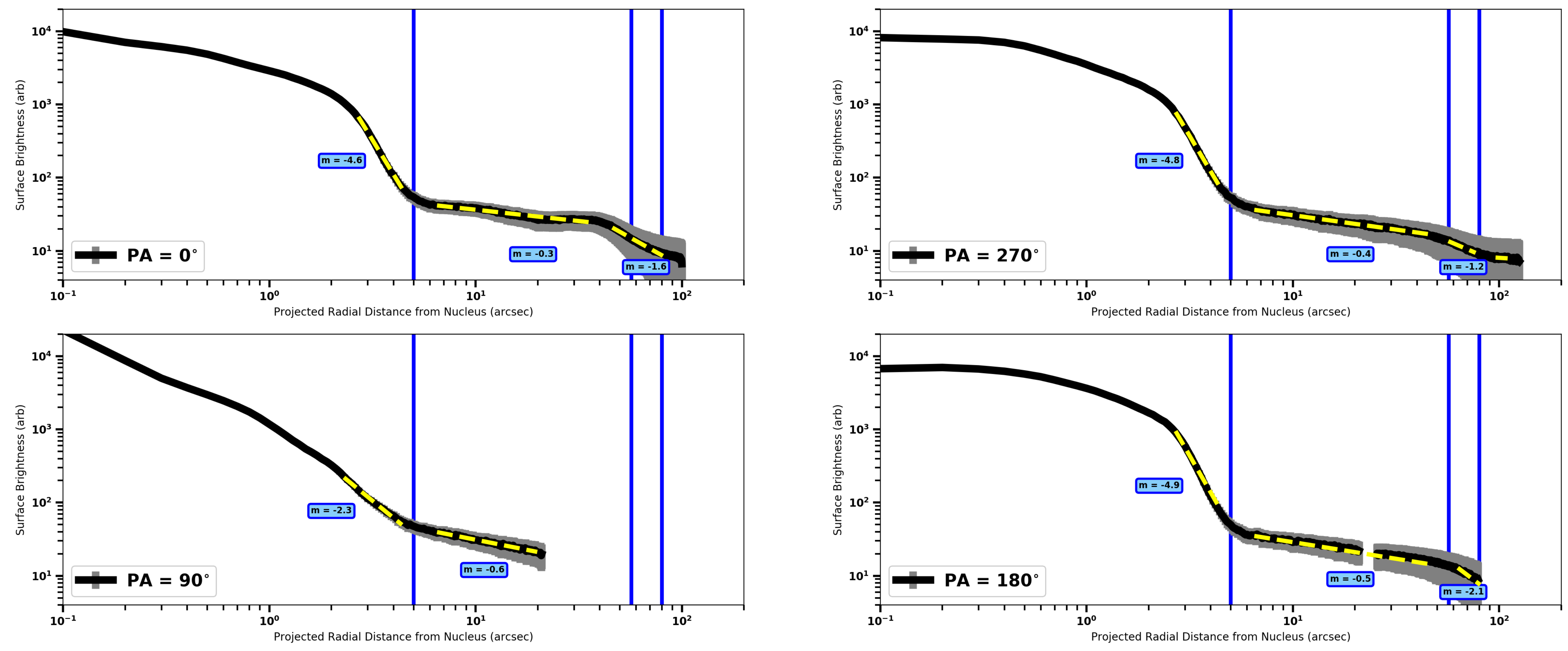}{1.0\textwidth}{}
          	}
\caption{Shown is an enlarged view of the radial surface brightness profiles highlighting the complex slope behavior of each region of the multi-component coma and the slope behavior of each region. The profiles contain both radial and azimuthal deviations from the canonical 1/$\rho$ behavior. Vertical blue lines separate the four regions of the coma's profile and yellow dashed lines indicate a 1/$\rho^m$ profile fit to each region. \label{fig:WFPC2_Profiles_zoom}}
\end{figure}

The comae are represented by complex monotonically deceasing profiles, which can be divided into five regions: (1) a rapidly expanding outburst dust coma contained within $5''$ of the nucleus (Region 1), (2) a monotonically decreasing dust coma detected in Exp-1 with similar slope between 1$''.4$-5$''$ (Region II), (3) a second monotonically decreasing dust coma detected in both Exp-1 and Exp-2 with similar slope between 5$''$-57$''$ (Region III), (4) a third monotonically decreasing dust coma detected in both Exp-1 and Exp-2 exterior to 57$''$ (Region IV), and (5) a background noise limited portion of the profile beyond $\sim$80$''$ (Region V). The most apparent feature of the surface brightness profiles is the similarity between both epochs of imaging beyond $5''$. The profiles are identical within uncertainties outside of $5''$ and indicate no detectable evolution of morphology during the 19.3 hours between imaging epochs. The recent outburst coma is fully contained within a $5''$ radius circular region centered on the nucleus' position. Region III is dominated by the residual expanding dust from an earlier 1996 February 17 outburst (\citet{2016Icar..272..327M}). A 20$''\times$10$''$ region of enhanced surface brightness is located near a 0$^{\circ}$ PA and is the result of either a higher density of dust grains and/or emission of dust grains characterized by a different particle size distribution emitted at this PA during the February outburst, resulting in a larger effective scattering cross-section. The final region of detectable comae is Region IV which approximately follows the canonical $m=$-1 slope behavior. Figure \ref{fig:WFPC2_Profiles_zoom} shows an enlarged view of the profiles for PAs 0$^{\circ}$, 90$^{\circ}$, 180$^{\circ}$, and 270$^{\circ}$ showing each of the profile regions fit to a 1/$\rho^m$ profile. The profile's slopes in Regions I \& III deviates significantly from the canonical $m=$-1 slope behavior indicative of an isotropic and steady state dust emission.

\begin{figure}[ht!]
\gridline{
		\fig{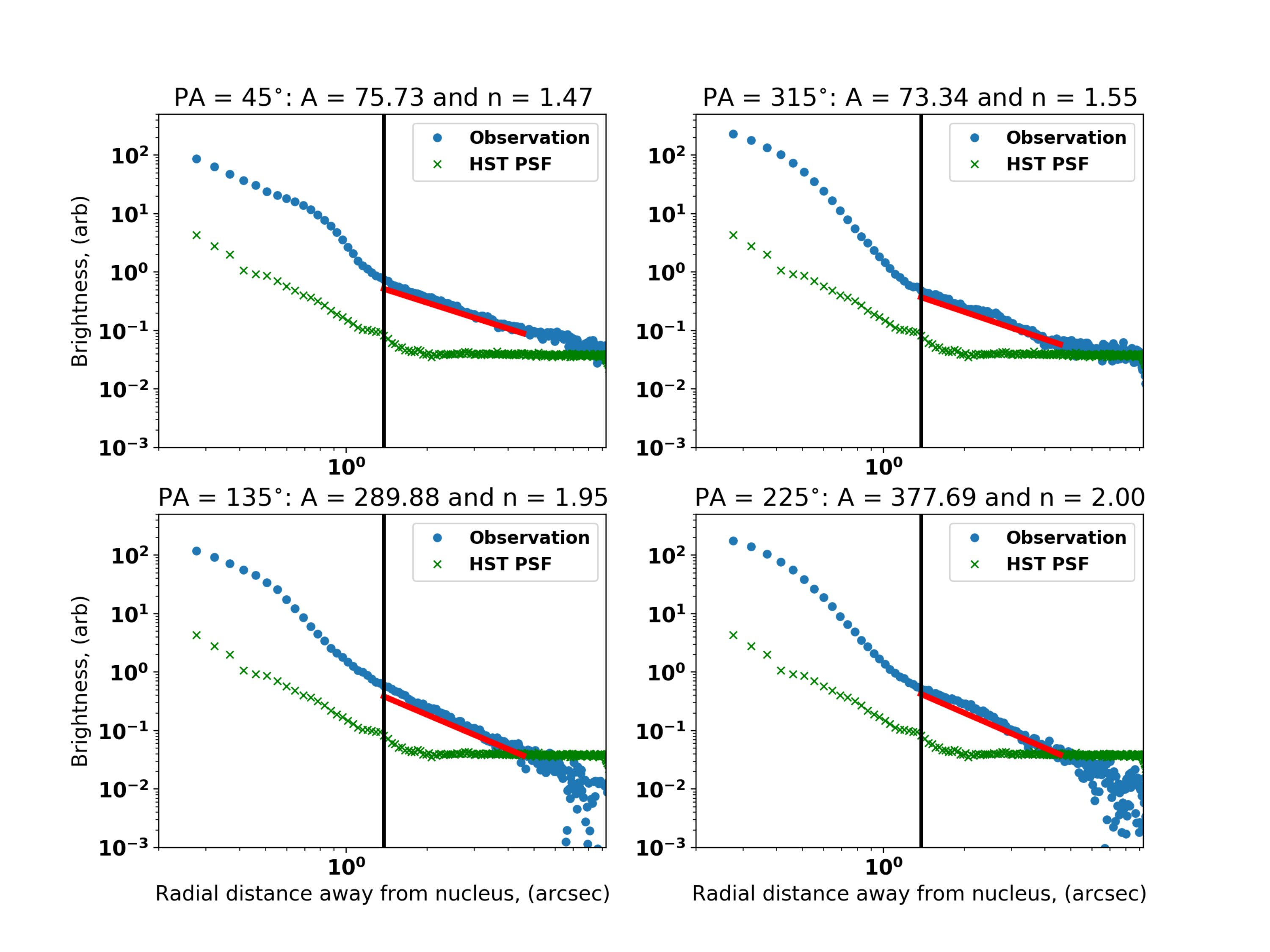}{0.75\textwidth}{}
          	}
\caption{Radial surface brightness profiles of Exp-1 for 20$^{\circ}$ pie-shaped wedges centered on the identified PAs. The most recent outburst coma is contained within a projected radius of 1$''$.4. and a quiescent coma is detectable outside of the outburst coma to a radial distance of $\sim$5-10$''$. The profiles between radii 1$''$.4 and 4$''$.6 were fitted with a profile of the form $A/\rho^n$. The red solid-line in each panel shows the results of these fits with the best-fit parameters indicated above each plot. The green profiles indicate the behavior of a PSF generated by Tiny Tim (\citet{2011SPIE.8127E..0JK}) combined with a modeled background level for the Exp-1 PC image. \label{fig:Exp-1_coma_slope}}
\end{figure}

The radial profiles in Figure \ref{fig:WFPC2_Profiles} and visual inspection of the Exp-1 PC images indicate the outburst dust coma detected in Exp-1 is contained within an $\sim$1$''$.4 radius aperture centered on the nucleus' position. Analysis of Region II (1$''$.4 $\leq \rho \leq$ 5$''$) in Exp-1 provides an opportunity to estimate the quiescent coma's flux contribution in Region II for the Exp-2 image, which is dominated by the most recent outburst dust coma. As will be investigated further in Section \ref{sec:photometry}, the 120 s Exp-2 image is the only observation which does not suffer from pixel saturation in the region surrounding the nucleus and provides the only opportunity for photometric measurements related to the most recent outburst dust coma (e.g. estimates of dust mass emitted during the outburst). To estimate the underlying quiescent coma's flux contribution within 5$''$ of the nucleus for the Exp-2 image we assume that the quiescent coma's behavior derived in Exp-1 is still present with minimal change in Exp-2. The radial surface brightness profiles in Exp-1 Region II were fit to a function of the form $A/\rho^n$. We chose to measure the Region II quiescent coma's profile behavior in four median-combined 20$^{\circ}$ pie-shaped wedges centered on PAs 45$^{\circ}$, 135$^{\circ}$, 225$^{\circ}$, and 315$^{\circ}$. Diffraction spikes present in the Exp-1 images were positioned approximately at PAs 0$^{\circ}$, 90$^{\circ}$, 180$^{\circ}$, and 270$^{\circ}$ (as seen more clearly in the enhanced images of Section \ref{sec:enhancements} Figure \ref{fig:enhanced_images}) and limited the range of suitable PAs for profile fitting, restricting profile characterization using a finer azimuthal spacing. Figure \ref{fig:Exp-1_coma_slope} shows the four profiles and results of the profile fitting. The Exp-1 derived quiescent coma's profile exhibited a north-south asymmetry and east-west symmetry. The final Exp-1 derived quiescent coma model used for further analysis was the result of averaging the two northern fit parameters and two southern fit parameters, producing a coma model with north-south asymmetry and east-west symmetry.  

\begin{figure}[ht!]
\gridline{
		\fig{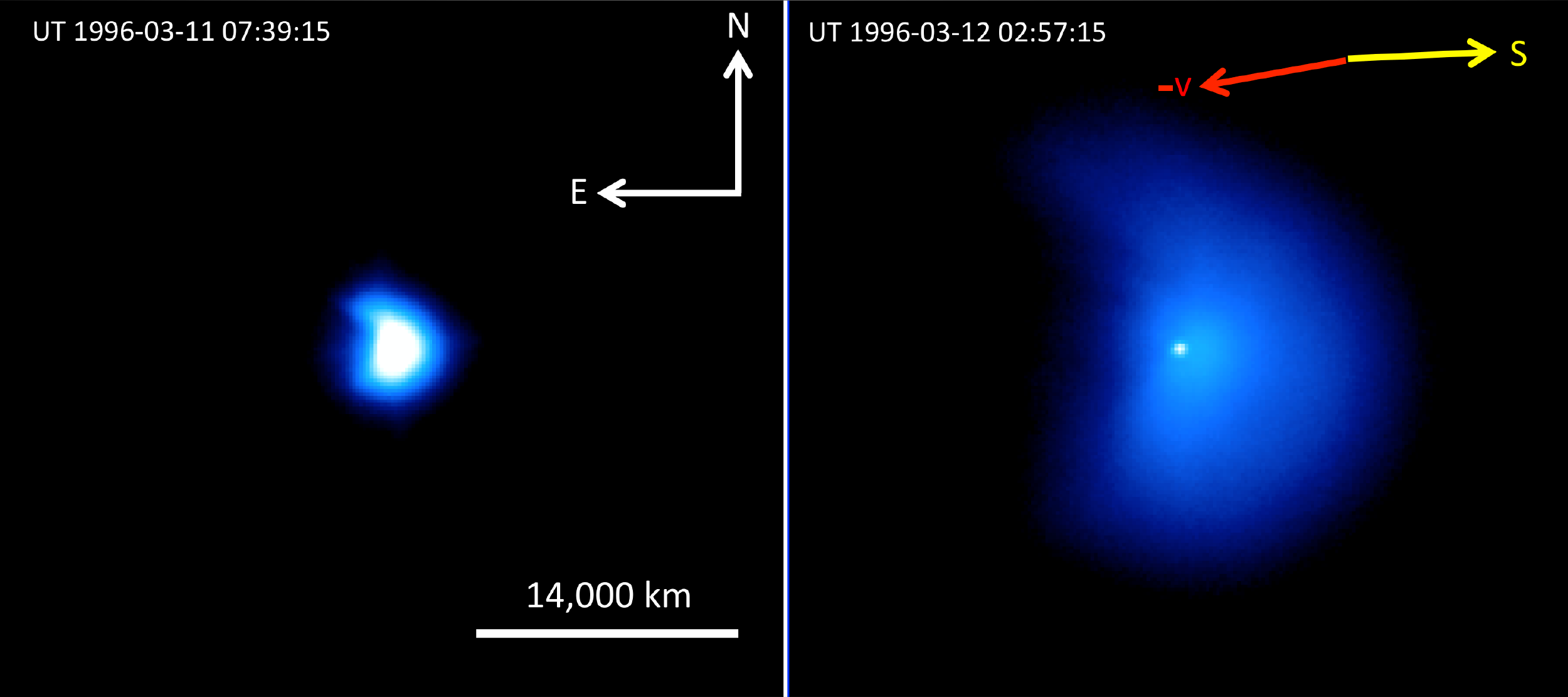}{1.0\textwidth}{}
          	}
\caption{Planetary Camera (PC) cropped observations of SW1 from 1996 observations. In both observations the arrows indicate the directions of equatorial coordinates. The position angles of the Sun and negative Centaur velocity vector are indicated by the yellow and red arrows. A scale bar on the bottom right of the Exp-1 image shows a projected distance of 14,000 km at the location of SW1 and is similar for Exp-2. \label{fig:PC_images}}
\end{figure}

The evolution of morphology detected in the most recent outburst coma contained within the 5$''$ radius aperture is ideally suited for analysis and 3-D Monte Carlo coma modeling efforts to place constraints on the underlying nucleus' spin state. In the remainder of this manuscript we focus on the analysis of this region. This outburst coma is fully contained within the higher-resolution PC's FOV, with a 0$''$.046/pixel native pixel scale. Because of this we proceed with our analysis using these higher-resolution data products instead of the mosaicked images shown in Figure \ref{fig:WFPC2}. Figure \ref{fig:PC_images} shows the 120 second Exp-1 and Exp-2 PC images of the outburst dust coma cropped to an 18$''.4$ square FOV. These 0$''$.046/pixel images, with an effective pixel scale of 176 km/pixel, combined with the WFPC2's stable F702W point spread function (PSF) of $\sim$$0''.06$ represent the highest resolution images of SW1's outburst coma to date. The outburst dust coma possesses a complex N-S symmetry and E-W asymmetry. This ``kidney'' shaped coma morphology is detected in both epochs of observations. So much so, Exp-2 resembles a scaled version of Exp-1, indicating dust grain emission during the outburst event only possessing a radial component of velocity and no angular momentum imparted from the rotating nucleus. The coma morphology contained in each observation is complex, showing radially curved features, however no appreciable signatures of rotation are noticeable through visual inspection of the raw images. The following subsection presents enhancements to these images that bring out low-level surface brightness variations.

\FloatBarrier

\subsection{Image Enhancements} \label{sec:enhancements}
\begin{figure}
\gridline{
		\fig{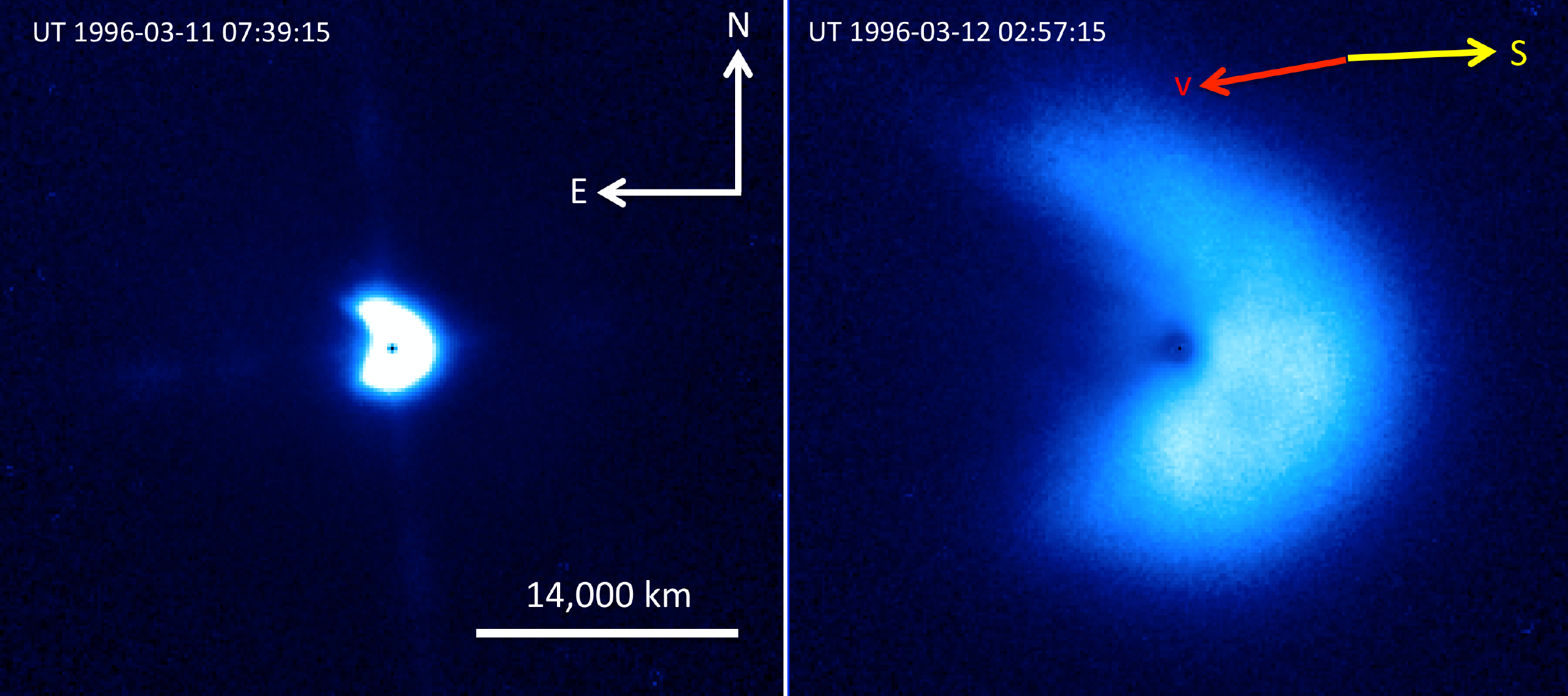}{0.5\textwidth}{(a)}
          	\fig{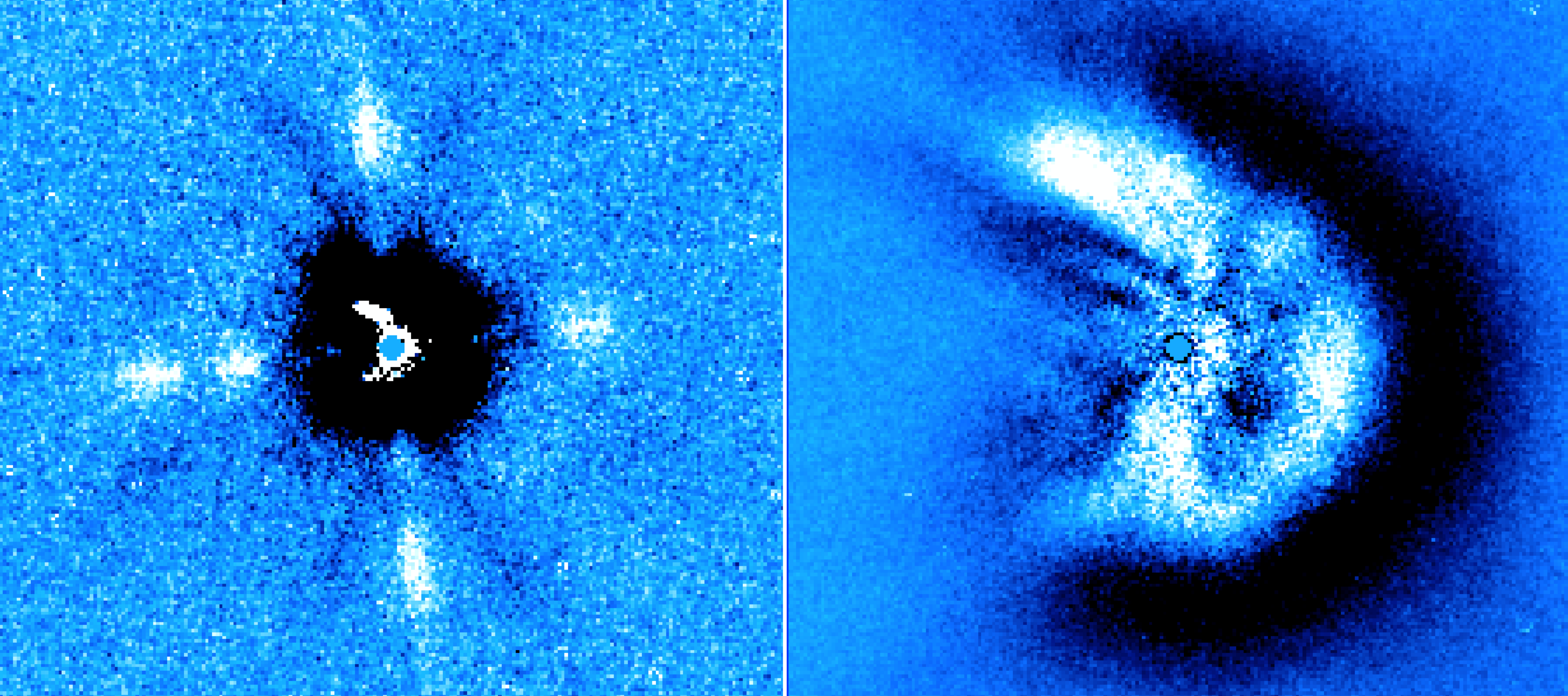}{0.5\textwidth}{(b)}
          	}
\gridline{
          	\fig{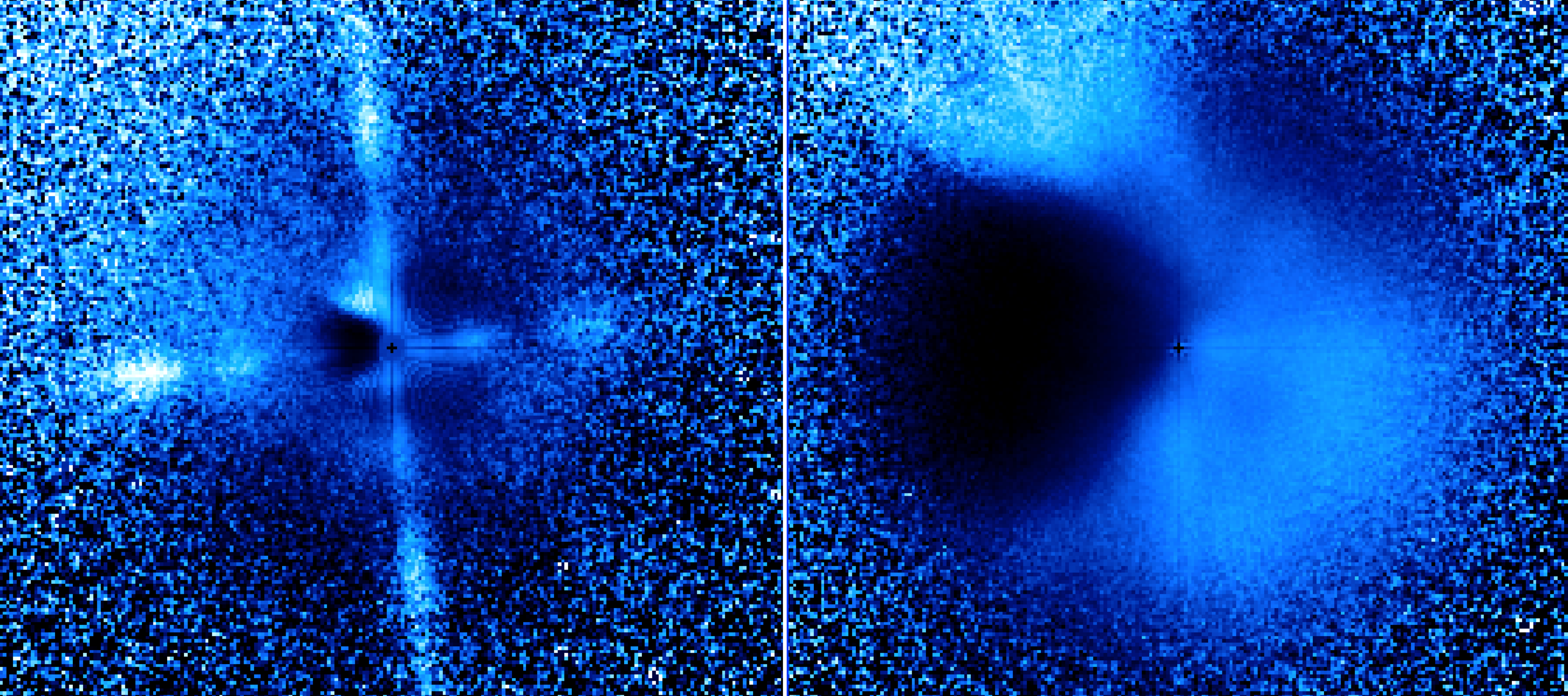}{0.5\textwidth}{(c)}
          	}
\caption{Panels showing the SW1 PC images (Exp-1 left \& Exp-2 right) after enhanced by: (a) division by a 1/$\rho$ profile, (b) application of a radially-varying spatial filter, and (c) division by an azimuthal median radial profile. While a complex morphology is seen in the enhanced images no features are present which suggest appreciable rotation of the nucleus while the dust grains composing the coma were being emitted. Exp-1 enhanced images in a, b, and c contain enhancements to the image's diffraction spikes.
\label{fig:enhanced_images}}
\end{figure}

The PC images were enhanced using techniques described in \citet{2014Icar..239..168S} to increase contrast of features present in the images. Figure \ref{fig:enhanced_images} shows the 120 second Exp-1 and Exp-2 PC images after enhancement routines have been applied, highlighting the outburst coma's complex structure. The enhancements shown for the 1996 HST observations are: division by a 1/$\rho$ profile (where $\rho$ is the projected distance from the nucleus), application of a radially-varying spatial filter (RVSF), and division by an azimuthal median radial profile.

Figure \ref{fig:enhanced_images}(a) displays the PC images after division by a 1/$\rho$ profile, enhancing the deviation of the outburst coma from an ideal steady state isotropic emission of dust grains. The asymmetric kidney shaped outburst coma is visible in both Exp-1 and Exp-2. Both coma morphologies lack surface brightness in a wedge shaped region between PAs 45$^{\circ}$-135$^{\circ}$ when compared to the western half of the outburst coma. Comparison of the residual surface brightness in this wedge region to the coma's brightness outside of the 5$''$ radius region suggest that no detectable dust grains were emitted into the wedge region during the outburst. The 1/$\rho$ enhanced Exp-2 image has regions of brighter surface brightness in the south-west quadrant of the image. This may be due to the higher cross-section of dust grains caused by a higher number density of grains in this region or grains that are larger than larger grains in other regions of the western side of the coma. Features detected in the RVSF enhanced Exp-1 and Exp-2 images shown in Figure \ref{fig:enhanced_images}(b) have similar morphologies, but possess moderate structural differences in the relative locations and level of small-scale structure seen in individual coma features. This is attributed to the inherent nature of the RVSF procedure selectively enhancing surface-brightness variations of features with different size scales differently (see Section 2.3(b) in \citet{2014Icar..239..168S}). The bright radial features in the Exp-1 enhanced image are due to the diffraction spikes present in Exp-1. One noticeable similarity between features is the curved feature originating near PA = 0$^{\circ}$ and extending to the east in both enhanced images. No evident rotational movement of PA for this feature is seen when comparing both images. The final enhancement technique we employed is division by an azimuthal median radial profile, which enhances azimuthal variations in the image (Figure \ref{fig:enhanced_images}(c)). These enhanced images indicate the same outburst coma morphology which was suggested by the 1/$\rho$ removed profiles and indicates no evolution of the features, aside from a radial expansion. The nature of the morphology detected in the enhanced images further suggests that the outburst coma's morphology during the two epochs of observations are scaled versions of each other and no detectible signatures of the nucleus' rotation are seen.

\subsection{Outflow Velocity and Outburst Time}
Measurements of the projected outflow velocity ($v_p$) for the outburst coma were accomplished by comparing feature positions of the radial surface brightness profiles between Exp-1 and Exp-2. The projected material outflow velocity is the product of the actual emitted dust grain's physical velocity ($\bf{v}$) and the angle this velocity makes with the sky plane. An azimuthal dependent projected outflow velocity was measured. The western half of the outburst coma (PAs between 180$^{\circ}$ and 360$^{\circ}$) had a consistent value of $v_p$ = 0.17 $\pm$ 0.02 km/s. Using this velocity and the position of the shell of material on the western half of the outburst coma, we derive an outburst date of UT 1996-03-10 23:38 $\pm$ 4 hours, 8 $\pm$ 4 hours before Exp-1. This value is consistent with estimates from an independent analysis of the HST images from \citet{2016Icar..272..327M}. Additionally, \citet{2016Icar..272..327M} discusses a ground-based image of SW1 acquired on UT 1996-03-10 21:59:40 (9.66 hours before Exp-1) that does not show signs of the outburst event, giving a constraint for the outburst time, which our derived outburst time is consistent with.  

The physical velocity of the dust grains emitted during the outburst were used to estimate the effects of solar radiation pressure on the morphology of the expanding outburst coma.  We calculated the distance $d$ traveled by a dust grain in the projected sky-plane of the observations before being turned back due to radiation pressure using Equation 2 from \citet{2013Icar..222..799M}:
\begin{equation}
d = \frac{\textrm{v}^2 \cos^2\gamma}{2 \beta g \sin{\alpha}}
\end{equation}

\noindent where $\textrm{v}$ is the magnitude of the initial physical ejection velocity of the dust grains, $\gamma$ is the angle between the initial direction of the dust grains and the sky-plane, $\beta$ is the ratio of the acceleration due to solar radiation pressure and solar gravity, which is inversly related to the size of the dust grains, $\alpha$ is the solar phase angle of the observations, and $g = GM_{\odot}/R^2_H$ is the gravitational acceleration on the dust grains. The initial physical velocity magnitude of the dust grains and their angle with respect to the sky-plane is not known a priori. In Section \ref{sec:monte_carlo} we show through 3-D Monte Carlo coma modeling that the best-fit 3-D shape of the outburst coma's western half is a cone of ejected material with 30$^{\circ}$ apex angle directed along the nucleus-Sun vector. This 3-D Monte Carlo coma modeling effort resulted in values of the initial physical velocity of the dust grains being $\textrm{v} = $ 0.25 $\pm$ 0.075 km/s and $\gamma$ = 75$^{\circ}$ for use in Equation 1. The dust grain size distribution of the outburst coma is not known, so we cannot directly determine the $\beta$ values present in the outburst coma. We instead calculated the turn-back distance for a range of plausible $\beta$ values (0.1, 0.6, 1.0) that are based on a quiescent activity SW1 {\it Spitzer} IRS spectrum derived dust grain size distribution characterized by a $dn/da \sim a^{-3.75}$ distribution (\cite{2015Icar..260...60S}). The $\beta$ values chosen represent the smallest dust grains contained in dust comae with similar grain size distributions (\cite{2004come.book..565F}) and represent the behavior of dust grains which would undergo the most deflection in coma morphology due to solar radiation pressure. Their associated projected turn-around distances are $d \approx$ (3,100,000 km, 516,000 km, and 310,000 km) for dust grains leaving the edge of the modeled source region. These distances are all well beyond the $5''$ radius region ($\sim$ 19,000 km) containing the outburst coma, indicating radiation pressure effects have not significantly altered the outburst coma's morphology.

\FloatBarrier
\subsection{Photometry}
\label{sec:photometry}
\begin{figure}
\gridline{
		\fig{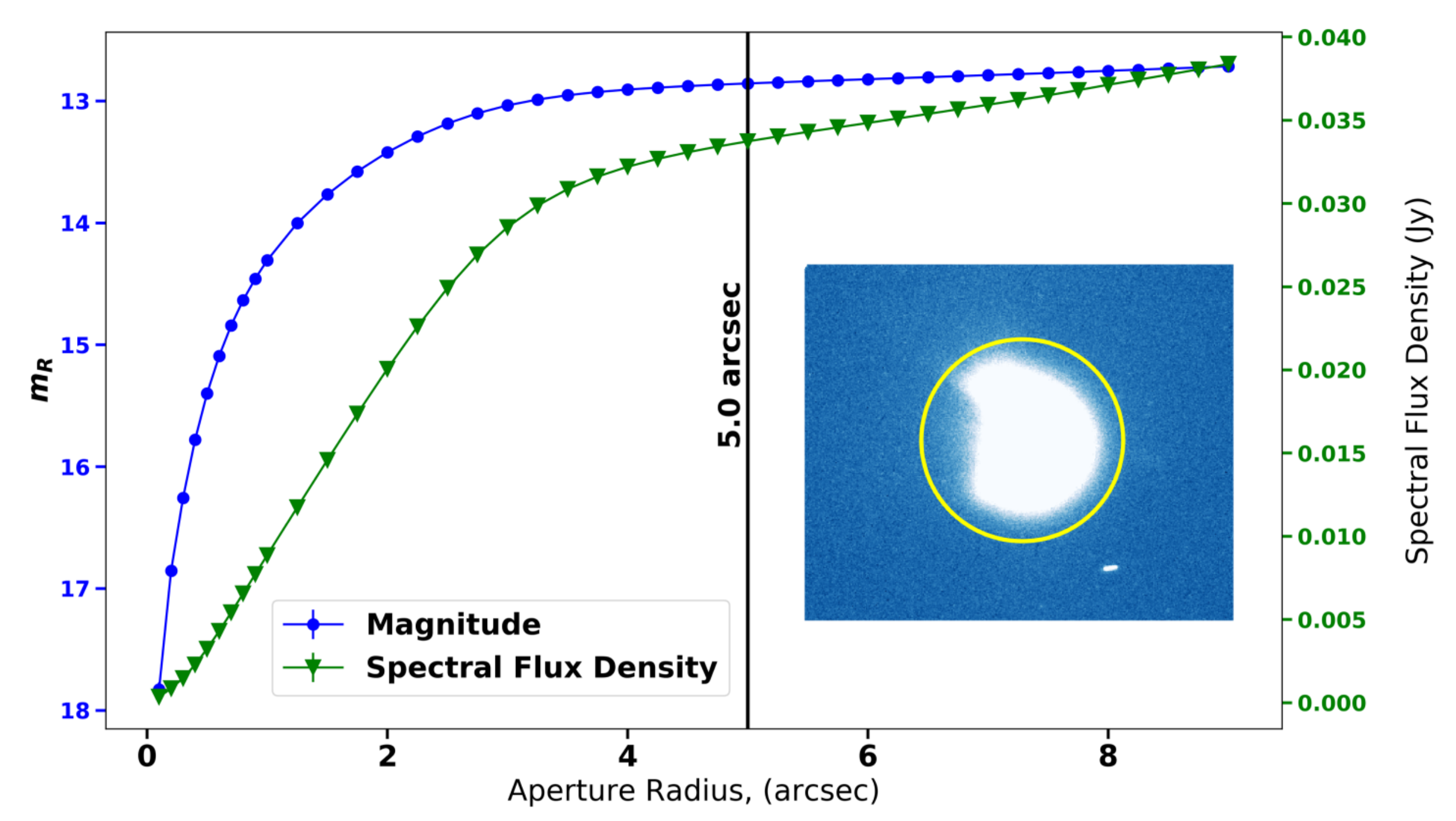}{1.0\textwidth}{}
          	}
\caption{Equivalent Cousins R-band magnitude measurements of the Exp-2 120 second PC observation for a series of apertures centered on the nucleus' location are shown by the blue dotted curve. Corresponding WFPC2 F702W-filter spectral flux density measurements are also indicated by the green triangles. Also included is a cropped PC image displayed using a zscale to highlight the dust emitted during the outburst as being contained within a 5$''$ radius aperture. A yellow circular line surrounding the outburst coma indicates the size of the $5''.0$ photometric aperture and the black vertical line indicated the position of $5''.0$ on the plot. \label{fig:photometry}}
\end{figure}

We used the 120 second Exp-2 PC image to measure SW1's brightness in a series of photometric apertures, following the procedures explained in the WFPC2 Data Handbook (\citet{2010hst...gonzaga}), to generate measurements in the STMAG system. The STMAG$_{\textrm{F702W}}$ measurements were then converted to equivalent Cousins R-band magnitudes using Equation 5 from \citet{1995PASP..107.1065H}:
\begin{equation}
m_R = \textrm{STMAG}_{\textrm{F702W}} + T_1 \times (m_V - m_R) + T_2 \times (m_V - m_R)^2 + Z
\end{equation}  

\noindent where $T_1$, $T_2$, and $Z$ are calibration conversion coefficients found in \citet{1995PASP..107.1065H} Table 3. The color term $(m_V - m_R) = 0.50 \pm 0.03$ was used and is the average of measurements from \citet{1993AJ....106.1222M} and \cite{2009AJ....137.4296J}, where both report consistent values for SW1 from observational epochs separated by 15 years. Figure \ref{fig:photometry} shows the equivalent $m_R$ measurements as well as spectral flux density measurements for a series of apertures centered on the nucleus' location. A $m_R (5''.0) = 12.86 \pm 0.02$ was measured for SW1 during the outburst. \citet{1996DPS....28.0808F} reported a ``R magnitude $\sim$ 12.5 in a 5$''$ radius'' inconsistent with our observed brightness, which could be the result of images from an earlier pipeline calibration process being used between each analysis. The photometry techniques used for the 5$''$.0 measurement of SW1 were checked on the trailed field star (J2000 coordinates: RA = 10:27:10.31 \& DEC = +04:20:46.85) in the PC image, which recovered the star's Pan-STARRS reported rMeanKronMag magnitude converted to an equivalent Cousins R-band magnitude ($m_R$ = 19.96), giving confidence in the methods used in this work.  

The photometric behavior of measurements made with increasingly larger photometric apertures (Figure \ref{fig:photometry}) reinforces evidence shown in Section \ref{sec:morphology}'s morphology argument that the majority of the scattered light from the outburst coma is contained within a 5$''$.0 radius aperture. We use this aperture to estimate a lower-limit on the total amount of dust emitted during the 1996 outburst using methods similar to our analysis of Kitt Peak 2.1-m observations of SW1 from 2008 while in outburst (\citet{2017Icar..284..359S}). The lower-limit of dust mass emitted during the outburst was measured using equations from Jewitt (1991) and \citet{2004come.book..223L}:
\begin{equation}
M_d = \frac{4 \rho_d a C_{d}}{3}
\label{eq:dust_mass}
\end{equation}

\noindent where, $a$ is the dust grain radius (assuming spherical dust grains of a single size), $\rho_d$ is the dust grain density, and $C_d$ is the effective total scattering cross section for the collection of dust grains measured from the flux contained in the 5$''$.0 aperture. While there is clearly a dust grain size distribution associated with the outburst, we assume a single 1-$\mu$m grain size for this calculation due to its efficiency at scattering light at R-band wavelengths. For our calculations we assumed a dust grain density of 1000 kg/m$^3$. The effective total scattering cross section of the dust grains is obtained by subtracting the nucleus' flux and Exp-1 derived quiescent-activity coma (see Section \ref{sec:morphology}) contributions from the 5$''$.0 aperture flux measurement. A value of 32.3 $\pm$ 3.1 km for the nucleus' effective radius (\citet{2019Icar..SW1}), a nucleus geometric albedo of 0.04, and a linear phase function of 0.04 mag/deg were used to calculate the nucleus' spectral flux density contribution that was subtracted from the measured spectral flux density of the 5$''$ radius aperture. $C_{d}$ is calculated using:
\begin{equation}
p_{d} \Phi_{d} (\alpha) C_{d} = 2.238 \times 10^{22} R_H^2 \Delta \pi 10^{0.4(m_{\odot} - m_d)}
\label{eq:scattering_cross}
\end{equation}

\noindent where, $p_d$ is the geometric albedo of the assumed 1-$\mu$m dust grain outburst coma, $\Phi_d (\alpha) = 10^{-0.4 \alpha \beta}$ is the phase function used for the dust grains, where a value of $\beta$ = 0.02 mag/deg was used for the dust grains (\citet{1987ApJ...317..992J}), $\alpha$ is the phase angle of the observations, $m_{\odot}=-27.09$ is the solar magnitude in the R-band, and $m_d$ is the R-band magnitude measured for the outburst dust from the coma's spectral flux density measurement. The geometric albedo of the assumed 1-$\mu$m outburst dust coma is not known. We use a low dust grain geometric albedo of 0.04 (\cite{2004come.book..577K}) based on SW1's quiescent-activity dust grain properties derived from analysis of a {\it Spitzer} IRS spectrum that indicated a dust coma abundant  in low albedo, micron-sized, and superheated (e.g. large $T$(dust)/$T$(blackbody)) amorphous carbon and amorphous Mg-Fe silicate grains (\cite{2015Icar..260...60S}). Both $R_H$ and $\Delta$ in the equation are expressed in units of au, which are stated in Table \ref{tab:geometry}. Using Equations \ref{eq:dust_mass} and \ref{eq:scattering_cross} and the outburst coma's spectral flux density measurement we calculate $(2.79\pm0.05) \times 10^8$ kg for an estimated lower limit for the emitted dust mass during the outburst.

\FloatBarrier
\section{Monte Carlo Coma Modeling}  \label{sec:monte_carlo}
To place constraints on the nucleus' spin state (i.e.  spin pole direction and spin period) we used a 3-D Monte Carlo coma model to simulate the most recent outburst coma contained in the PC images. Table \ref{tab:monte_carlo} describes the model's input parameters, which are categorized into known and free parameters. The known parameters include the observational geometry and the time between the two observational epochs. The heliocentric and HST-nucleus distances had minimal change between the two epochs of observations and thus the same values were used for modeling each epoch. Unknown model input parameters include: the right ascension and declination of the nucleus' angular momentum vector, the rotation period of the nucleus, the cometographic longitude and latitude of the source region on the nucleus' surface, the size of the source region, the angular dispersion of the emitted dust grains, the length of time between the onset of the outburst and the observational epochs, the initial dust grains' velocities, and the outburst duration.

\begin{deluxetable*}{ll}
\tablenum{2}
\label{tab:monte_carlo}
\tablecaption{3-D Monte Carlo Coma Model Input Parameters}
\tablewidth{0pt}
\tablehead{
\colhead{Known Parameters} & \colhead{Free Parameters}
}
\startdata
		Heliocentric distance (6.26 AU)										&	Right ascension and declination of angular momentum				\\
		HST-nucleus distance (5.29 AU) 									& 	\hspace*{0.2cm} vector ($\alpha_{\textrm{c}}$, $\delta_{\textrm{c}}$)		\\
		Right ascension and declination of the Sun$^{\textrm{a}}$ 				&	Rotation period of nucleus (P)									\\
		\hspace*{0.2cm} Epoch 1: (RA =339.225$^{\circ}$ , DEC = -04.215$^{\circ}$) 	&	Cometographic longitude and latitude of source center 				\\
		\hspace*{0.2cm} Epoch 2: (RA =339.267$^{\circ}$ , DEC = -04.191$^{\circ}$) 	& 	\hspace*{0.2cm} ($\phi_{\textrm{s}}$, $\lambda_{\textrm{s}}$)			\\
		Right ascension and declination of the Earth$^{\textrm{a}}$				&	Size of the active region on the nucleus ($r_a$)						\\
		\hspace*{0.2cm} Epoch 1: (RA = 336.888$^{\circ}$, DEC = -04.318$^{\circ}$)	& 	Angular dispersion of emitted dust velocity ($\theta_d$)$^{\textrm{b}}$	\\
		\hspace*{0.2cm} Epoch 2: (RA = 336.800$^{\circ}$, DEC = -04.345$^{\circ}$)	& 	Time between outburst onset and first observations ($\Delta t_1$)		\\
		Time between two Epochs of Observations (19.3 hours)					&	Initial physical velocity of dust grains ($\textrm{v}$)								\\
																	& 	Outburst duration ($\Delta t_2$)								\\
\enddata
\tablecomments{\\
$^{\textrm{a}}$ RA and DEC as viewed from SW1 at the time of the observations.  \\
$^{\textrm{b}}$ Angular dispersion of the initial dust grain velocity direction, with 0$^{\circ}$ being normal to the local surface plane.
}
\end{deluxetable*}

The final synthetic SW1 images, which were compared to the two epochs of HST observations, are the result of combining the 3-D Monte Carlo coma model, a point source representing the nucleus' flux contribution, the Exp-1 derived quiescent-activity coma model, and an image-noise model derived from the noise behavior measured in Exp-1 and Exp-2 observations. The nucleus' point source was modeled by a HST point spread function (PSF) generated by the Tiny Tim software (\citet{2011SPIE.8127E..0JK}). The modeled outburst coma was convolved with the PSF to replicate the behavior of HST's imaging of the coma's extended emission. Generating the synthetic observations involved a linear combination of the Tiny Tim generated PSF, the 3-D Monte Carlo coma model, the Exp-1 derived quiescent-activity coma, and the background noise models. The linear scaling coefficients for the PSF and modeled outburst coma were determined by using photometry of the one Exp-2 image which did not contain saturated pixels. To determine the scaling factors we required photometric measurements for two apertures on the synthetic images to match measurements from the Exp-2 observations. The first aperture was centered on the nucleus and circular with 0$''$.22 radius. The second was a semi-circular annular aperture with 0$''$.5 inner and 0$''$.55 outer radii, centered on the nucleus, and restricted to the western half of the coma by using only regions between position angles 180$^{\circ}$ to 360$^{\circ}$. The circular region contained flux contributions from both the nucleus and outburst coma while the semi-circular annular region contained only a significant flux contribution from the outburst coma. The photometric measurements resulted in a linear system of two equations and with two unknowns that were solved to determine the scaling coefficients. Due to the saturation of pixels in regions surrounding the nucleus for Exp-1 we could not apply that same approach to scaling this synthetic observations. Instead, we used the nucleus' flux value derived from the Exp-2 analysis and only performed the photometric matching for the annular aperture to scale the model coma. The Exp-1 derived quiescent-activity coma and noise models were derived directly from the observations and as such were already scaled to appropriate flux levels.

For our Monte Carlo coma modeling efforts we only considered principal-axis rotation states due to the nucleus' large effective size and the short damping timescales for an $R_N\sim$30 km radius nucleus. We calculated the damping timescale using methods summarized in \citet{2008M&PS...43.1063S} and assuming standard cometary-like nucleus material properties. The damping timescale of a nucleus in an excited non-principal-axis state of rotation can be estimated by 
\begin{equation}
\tau_{\textrm{damp}} = \frac{\textrm{K}_1 \mu \textrm{Q}}{ \rho \textrm{R}^2_{\textrm{N}} \Omega^3}
\label{eq:damping}
\end{equation}

\noindent where K$_1$ is a non-dimensional scaling coefficient, $\mu$ is a parameter representing the rigidity of the nucleus material, Q represents a ``quality" factor for the nucleus material, $\rho$ is the bulk density of the nucleus, R$^2_{\textrm{N}}$ is the radius of the nucleus, and $\Omega$ is the angular velocity of the nucleus. \citet{2008M&PS...43.1063S} summarizes the uncertainties in our current knowledge of the material properties described by the parameters K$_1$ and the often-combined parameter $\mu$Q. We chose values of these properties believed to be reflective of the bulk material properties of JFCs nuclei: K$_1$ = 10 and $\mu$Q = 10$^7$ dyne cm$^{-2}$ (\citet{2008M&PS...43.1063S}). The bulk density of SW1's nucleus is not known, so we employed an estimated average for the bulk density for an ensemble of JFCs density estimates: $\rho$ = 600 kg m$^{-3}$ (\citet{2019SSRv..215...29G}). Since the angular velocity of the nucleus is not known a priori we calculate the damping timescale for a range of plausible assumed nucleus spin periods P = (0.5, 1.5, 3.0, and 60) days. These resulted in the following nucleus damping timescales: 0.17, 4.5, 36.2, 290,000 years. If SW1's spin period is in the range of typical JFCs nuclei spin periods between $\sim$3 to 40 hours (\citet{2017MNRAS.471.2974K}) SW1's large nucleus size would insure damping to a principal-axis rotation state after only a few years. If SW1's nucleus is in a slowly rotating state with spin period on the order of tens of days the assumption of principal-axis rotation breaks down and we can't rule out non-principal axis rotation. For the observational baseline of 19.3 hours between the two epochs of HST SW1 imaging this non-principal rotation state possibility does not impact the results of an analysis assuming principal-axis rotation. The 19.3 hour observational baseline is a small fraction of the rotational phase of an P = 60 day rotating body and the impact for our non-principal axis rotation state assumption is negligible.

Initially the distinctive equatorial east-west asymmetry and north-south symmetry of the outburst coma's morphology were considered to be primarily the result of the nucleus' rotation. However, results of a 3-D Monte Carlo coma modeling effort utilizing possible spin-pole orientations covering the full 4$\pi$ steradians of parameter space using 25$^{\circ}$ angular steps in right ascension and declination and spin periods ranging on the order of hours to days aimed at replicating the observed asymmetric morphology through nucleus rotation did not return any plausible spin states which were able to generate similar morphology in synthetic images. The observed asymmetric kidney shaped morphology appears instead to be the result of a non-isotropic emission of dust grains during a single, relatively short in duration when compared to the nucleus' rotation period, outburst event. 

\begin{figure}
\gridline{
		\fig{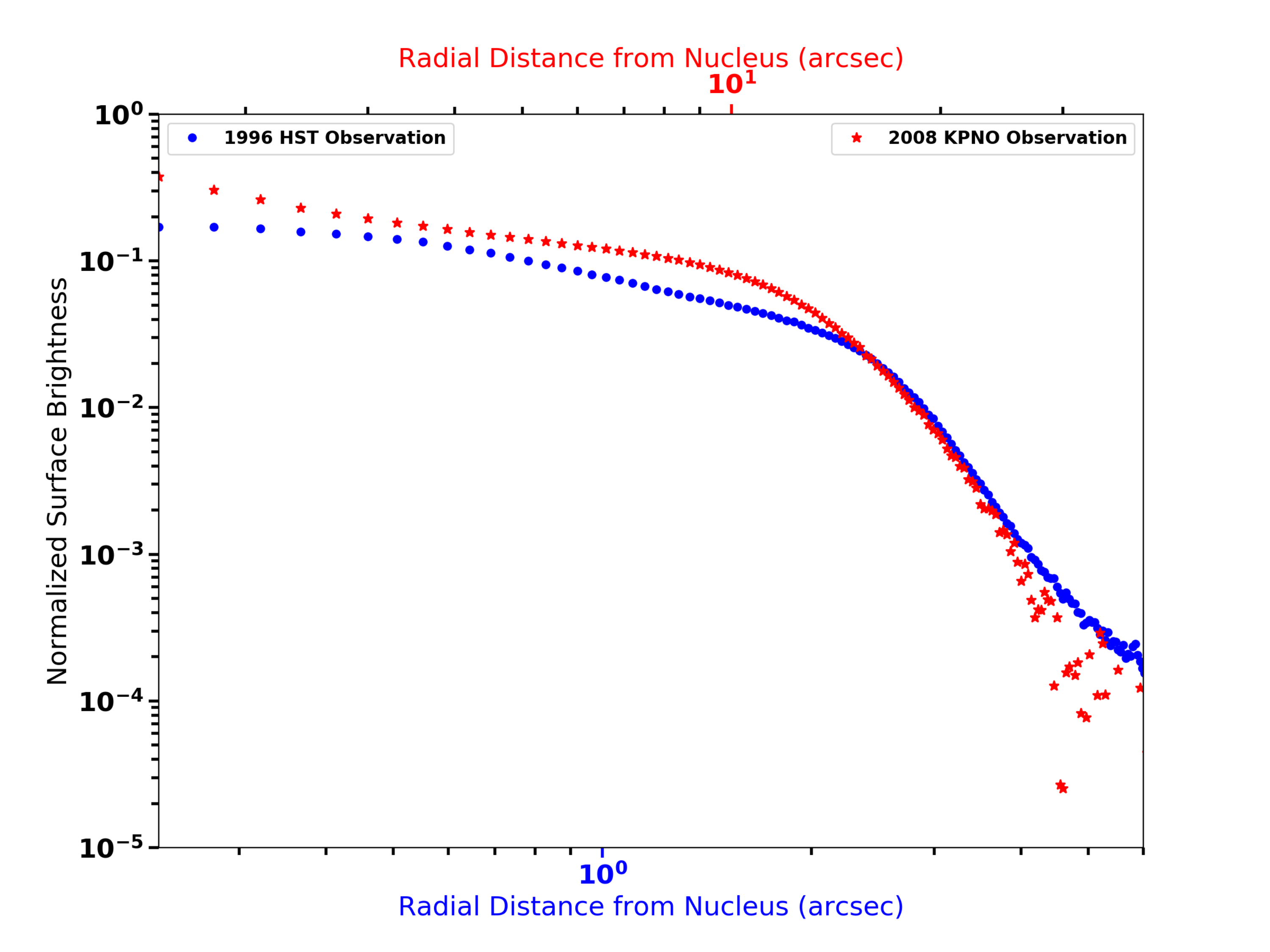}{1.0\textwidth}{}
          	}
\caption{Comparison of radial surface brightness profiles of the HST Exp-2 and 2008 Kitt Peak 2.1-m outburst observations. The radial surface brightness profiles represent the median surface brightness profiles for the western half of both images. The projected distances of the observations have been scaled to highlight similarities in the morphology at different spatial scales of the outburst comae's evolution. The scale on the bottom of the plot is for the HST image and the scale on the top is for the 2008 Kitt Peak image. \label{fig:2008_vs_1996}}
\end{figure}

Visual inspection of the western half of the outburst coma and comparison with an earlier analysis by our group of a set of five consecutive nights of Kitt Peak 2.1-m observations of a 2008 outburst of SW1 (\citet{2017Icar..284..359S}) suggests a similar 3-D shape evolution of the dust grains which were lofted from the surface by the short-lived outburst event. The outburst coma morphology detected and its evolution during five consecutive nights of imaging in 2008 were modeled using the same 3-D Monte Carlo code with the ``best-fit" model being represented by an expanding semi-symmetric cone of material directed along the nucleus-Sun vector. Excluding the eastern half of the 1996 outburst coma, the 1996 and 2008 outburst observations both possess a nearly identical expansion behavior as detected in the outburst comae imaging. The images visually have nearly identical morphology. Figure \ref{fig:2008_vs_1996} shows a comparison between radial surface brightness profiles for the 1996 HST Exp-2 and the first night of five consecutive nights of 2008 Kitt Peak 2.1-m imaging. For reference the Exp-2 image was obtained $\sim$25 hours after the start of the 1996 outburst event and the 2008 Kitt Peak image was obtained $\sim$ 114 hours after the start of the 2008 outburst event (an approximately five-times longer interval between outburst event and observational epoch when compared with the 1996 HST Exp-2 image). In Figure \ref{fig:2008_vs_1996}, the radial scale for the 2008 Kitt Peak profile has been set to $\sim$5 times the HST Exp-2 radial profile, which provides considerable overlap between the profiles. The behaviors of the surface brightness profiles are strikingly similar. Because of this similarity, combined with the inability to replicate the observed coma morphology seen in the 1996 HST observations from nucleus rotation, we proceeded with modeling the coma's morphology by assuming the outburst is best represented by a single outburst event lofting dust from a single source region located near the sub-solar point on the nucleus' surface. Most importantly, the outburst coma's asymmetric 3-D shape and the resulting coma morphology seen in the observations are the result of a cone of material directed along the nucleus-Sun direction. The source of the asymmetric, complex coma morphology is unknown, but is not unique to this 1996 HST epoch of observations. The recent works \cite{2016Icar..272..387M} and \cite{2016Icar..272..327M} give a thorough description of SW1's frequently occurring complex outburst coma morphology. A detailed analysis focused on understanding the nucleus processes generating the observed morphology is beyond the scope of this current work and will be the focus of a future manuscript. What is important for our current analysis is that the outburst coma morphology results from a single, impulsive, and short lived (relative to the nucleus' rotation period) event and we proceed with placing constraints on the minimal rotation period of the nucleus that can replicate the observed coma morphology.

The first step in our analysis was to determine the 3-D coma's structure which best replicated the observed morphology using a nucleus which has a rotational displacement which is negligible during the outburst duration. For this we chose a rotation period P = 10,000 days and an outburst duration $\Delta t_2$= 10.5 hours. This outburst duration is the middle value for a range of limits constrained from our earlier work (\citet{2017Icar..284..359S}) and was used for all further modeling efforts in this analysis. The exact value of the outburst duration is unknown and our analysis to place constraints on the nucleus' spin period is really placing constraints on the ratio of spin period to outburst duration: P/$\Delta t_2$. Recent observational constraints by \cite{2018EPSC...12..523M} have provided a lower limit to the outburst duration for an event with a total $\Delta m=2.0$ magnitude change being on the order of $\sim 1-2$ hours. This derived outburst duration is based on the time for 5$''$ radius aperture photometry measurements to rise from a quiescent activity level of 15.7 to a maximum value of $\sim$13.5. It is unknown whether SW1's outbursts with different total magnitude variations ($\Delta m$) correspond to different outburst durations (e.g. larger $\Delta m$ implying longer $\Delta t_2$). The magnitude change associated with the 1996 HST outburst being on the order of $\Delta m \sim$3.4 (based on observations reported in \cite{2016Icar..272..327M} by H. Mikuz from UT 1996 March 10.92 which captured SW1's magnitude in a 5$''$.8 radius aperture at 16.24 $\pm 0.05$) may have a slightly longer outburst duration required to loft the increased number of dust grains associated with the larger event. To help disentangle the relationship between outburst duration and outburst magnitude more information on the outburst mechanism(s) must be known. For the current work, we emphasize that our 3-D Monte Carlo coma modeling efforts are placing constraints on the ratio of the nucleus' spin period to outburst duration.

\begin{figure}
\gridline{
		\fig{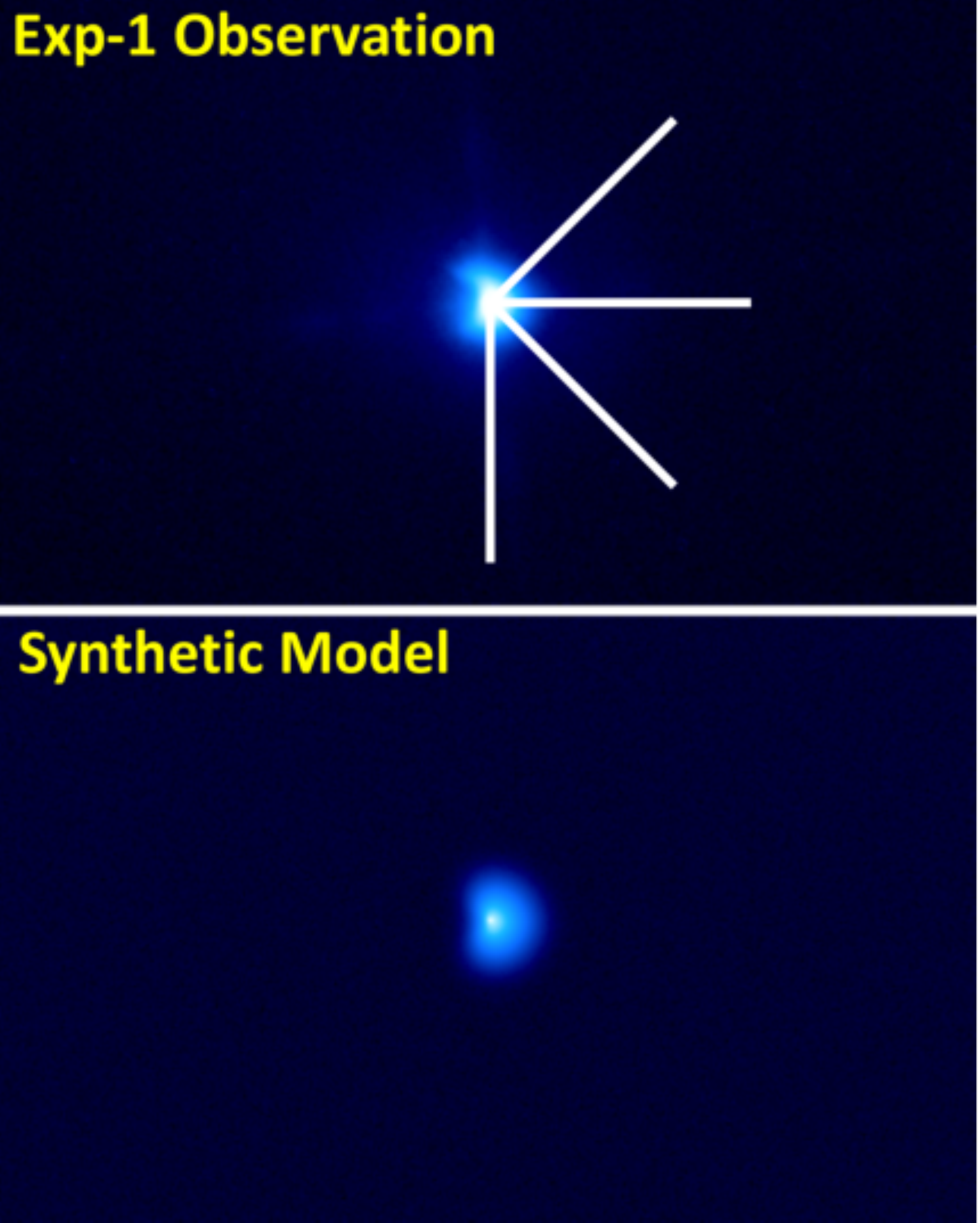}{0.275\textwidth}{(a)}
          	\fig{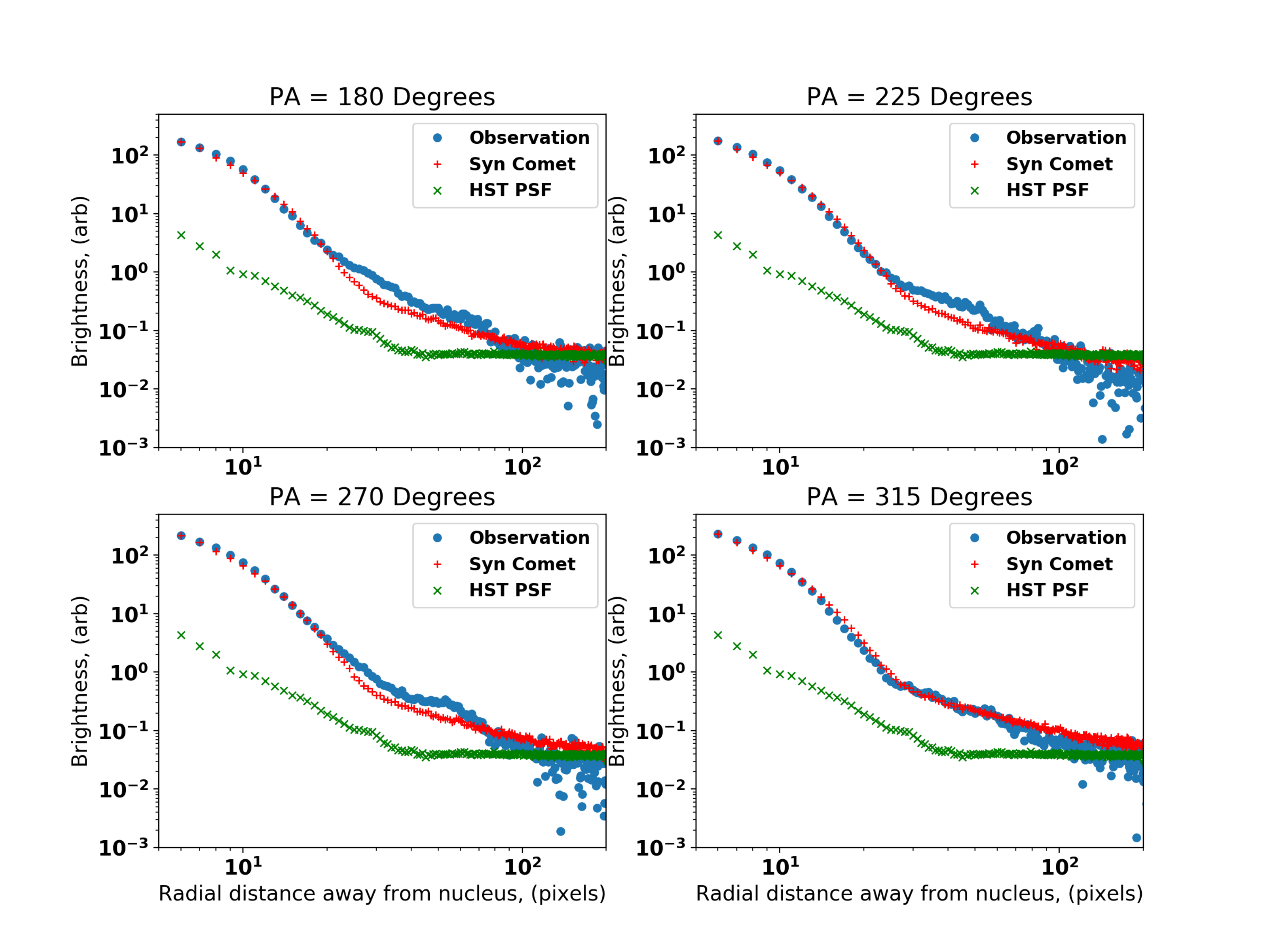}{0.5\textwidth}{(b)}
          	}
\gridline{
		\fig{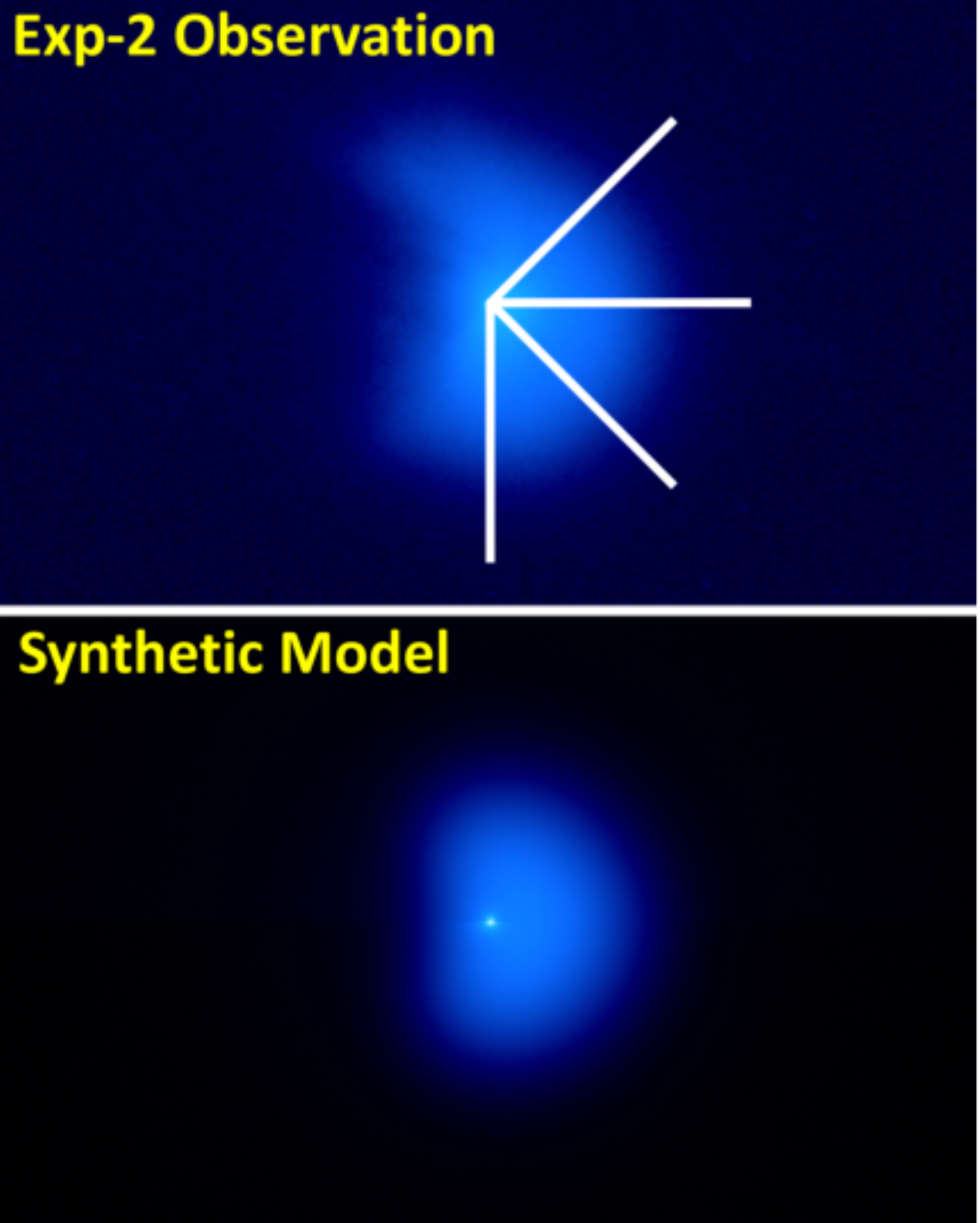}{0.275\textwidth}{(c)}
          	\fig{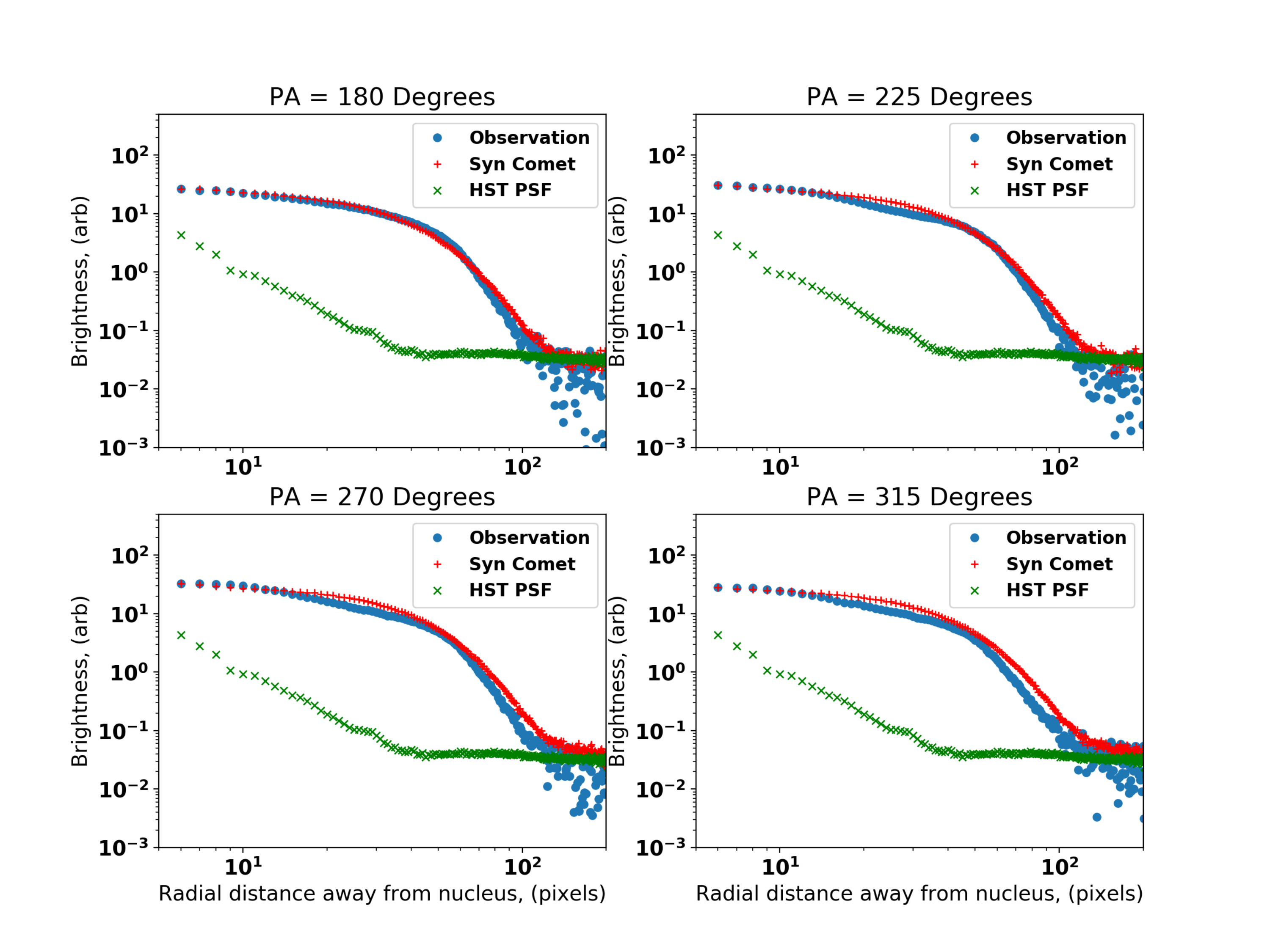}{0.5\textwidth}{(d)}
          	}
\caption{Panel showing the best-fit 3-D Monte Carlo outburst coma model compared with the observations. (a) Panel showing the observation (top) and synthetic image (bottom) for Exp-1. The four white lines on the observations indicate the PAs of the radial profiles shown in (b). The asymmetric nature of the outburst coma was replicated using the cone of material 3-D structure, with material suppressed from emission on the eastern half of the coma. (b) Radial surface brightness profiles of the observation, synthetic model, and Tiny Tim PSF. The overall structure of the complex coma was replicated well. Panels (c) and (d) are similar to (a) and (b) but are for the Exp-2 epoch of observations.  \label{fig:cone_outburst}}
\end{figure}

To quantitatively compare the synthetic models to the observations we calculated the weighted difference between the radial surface brightness profiles for position angles using an azimuthal spacing of 10$^{\circ}$. To calculate the weighted difference between the surface brightness profiles we constructed radial surface brightness profiles for 10$^{\circ}$ angular wedges centered on the nucleus' position that represent the median pixel value contained within the angular wedge. The weighted difference between profiles for each position angle and radial position were computed with $ (N_o - N_s)  / N_o$, where $N_o$ represents the pixel values for the HST observations and $N_s$ represents the pixel values for the synthetic outburst image. The best-fit model for this configuration is shown in Figure \ref{fig:cone_outburst}(a) and (c). The equatorial east-west asymmetric nature of the outburst coma morphology was replicated by suppressing dust grain emission for a portion of the eastern-half of the cone of material. Figure \ref{fig:cone_outburst}(b) and (d) show examples of the radial surface brightness profile comparison for four position angles on the western half of the outburst coma for Exp-1 and Exp-2. The 3-D outburst coma shape that best replicated the morphology detected in observations is a cone of material with conic apex (vertex) angle ($30 \pm 5)^{\circ}$, with a $\textrm{v}$= 0.25 $\pm$ 0.075 km/s initial physical dust grain velocity directed radially away from the nucleus, and the outburst taking place $\Delta t_1$= 8 $\pm 4$ hours before the Exp-1 observations. Uncertainties in model input parameters were constrained by determining the spread in input parameter space for a specific variable that resulted in a 1-$\sigma$ spread in the weighted difference between the observations and the synthetic models.

\begin{figure}
\gridline{
		\fig{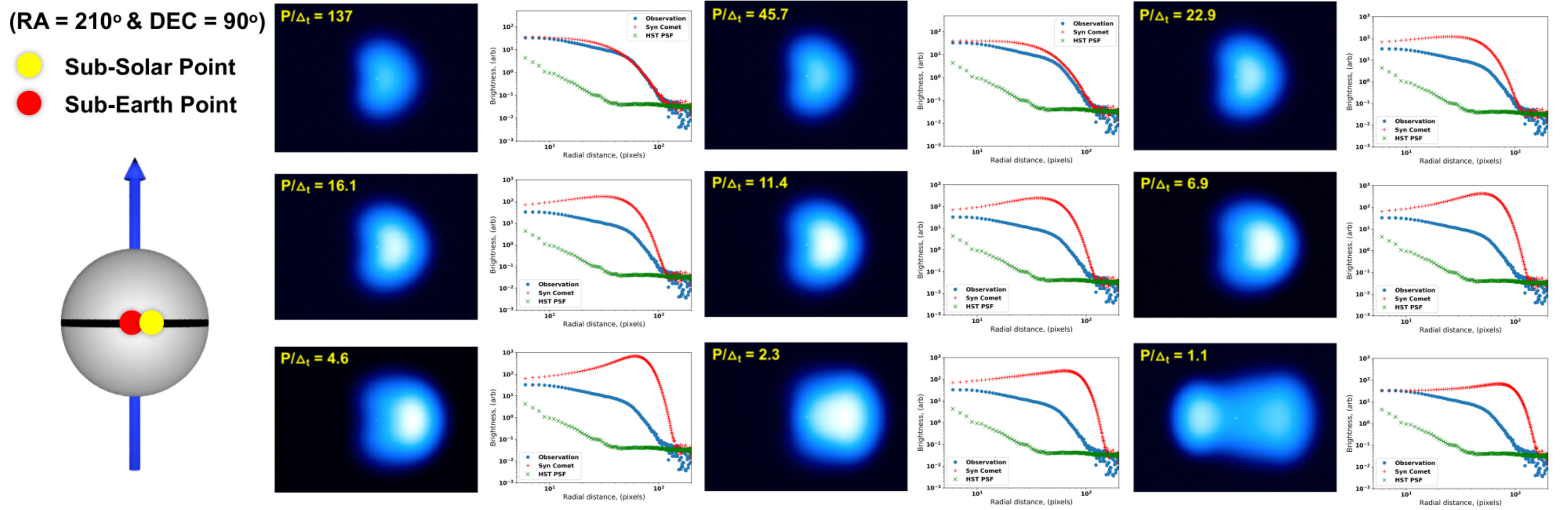}{0.9\textwidth}{}
          	}
\caption{Monte Carlo modeling results for a nucleus spin-pole orientation assumed to be perpendicular to the plane generated by the nucleus-Sun and nucleus-HST vectors. The diagram on the left depicts a notional spherical nucleus, with the blue arrow indicating the assumed spin-pole direction. The black line represents the nucleus' equator and red and yellow circles indicates sub-Earth and sub-solar locations on the surface. Synthetic images for the Exp-2 epoch and radial surface brightness profiles for a PA = 270$^{\circ}$ are shown to the right for comparisons between observations and synthetic model images. The orientations of the synthetic images are the same as in Figure \ref{fig:PC_images}. The results are expressed in terms of the ratio of spin period (P) to outburst duration ($\Delta t_2$): (left to right and top to bottom: 137, 45.7, 22.9, 16.1, 11.4, 6.9, 4.6, 2.3, and 1.1). For this assumed spin-pole orientation deviations between the observations and synthetic images are apparent for P/$\Delta t_2$ $\leq$ $\sim$40. \label{fig:spin_pole_1}}
\end{figure}

\begin{figure}
\gridline{
		\fig{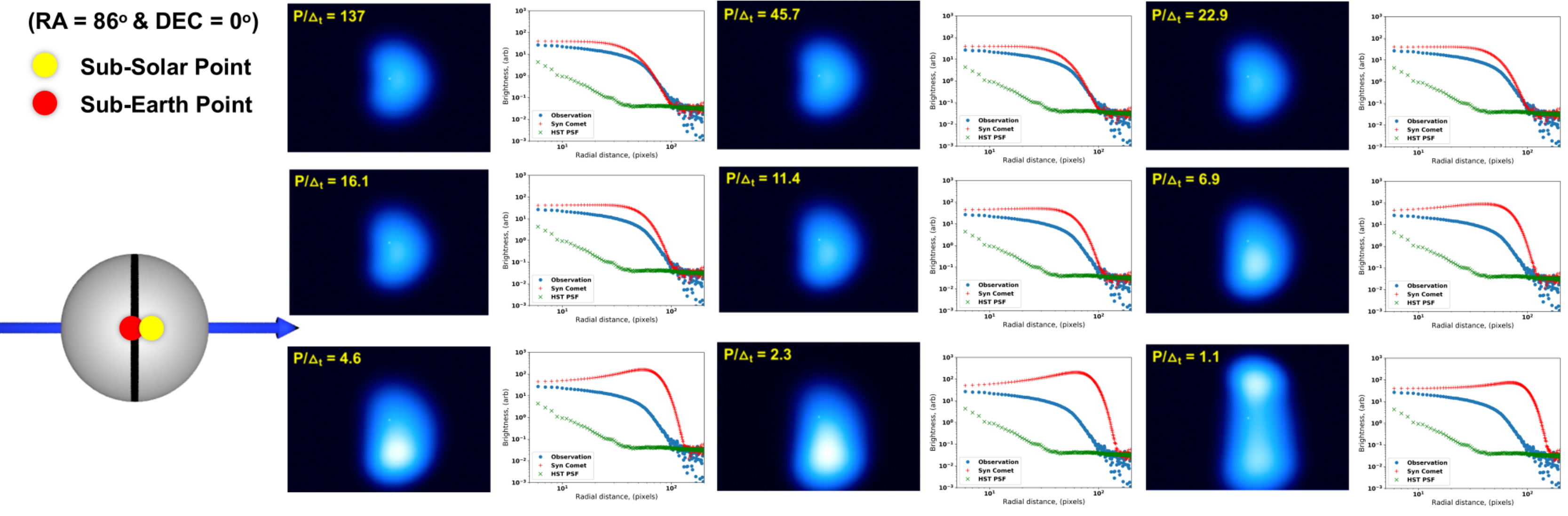}{0.9\textwidth}{}
          	}
\caption{Diagram is similar to Figure \ref{fig:spin_pole_1}, but for an assumed nucleus spin-pole direction indicated at the left. This assumed spin-pole orientation is parallel to the skyplane of the observations and along the projected direction of equatorial west. The radial profiles shown on the left are for a PA = 180$^{\circ}$. For this assumed spin-pole orientation, deviations between observations and synthetic images are apparent for P/$\Delta t_2$ $\leq$ $\sim$20.  \label{fig:spin_pole_2}}
\end{figure}

\begin{figure}
\gridline{
		\fig{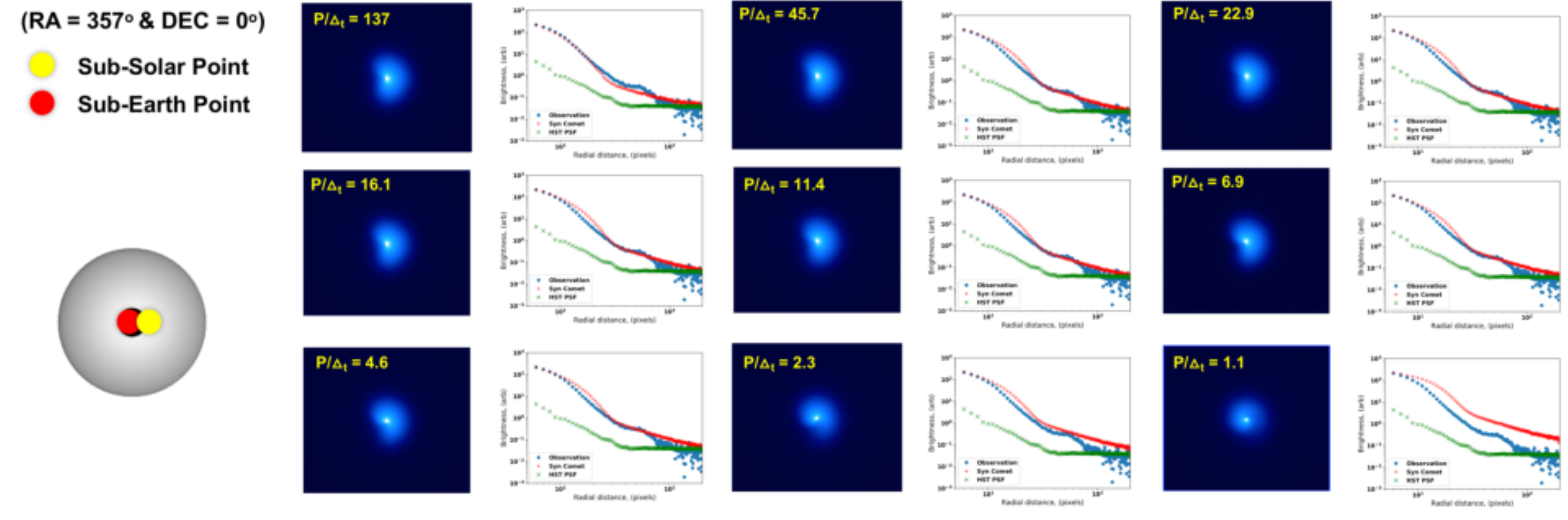}{0.9\textwidth}{}
          	}
\gridline{
		\fig{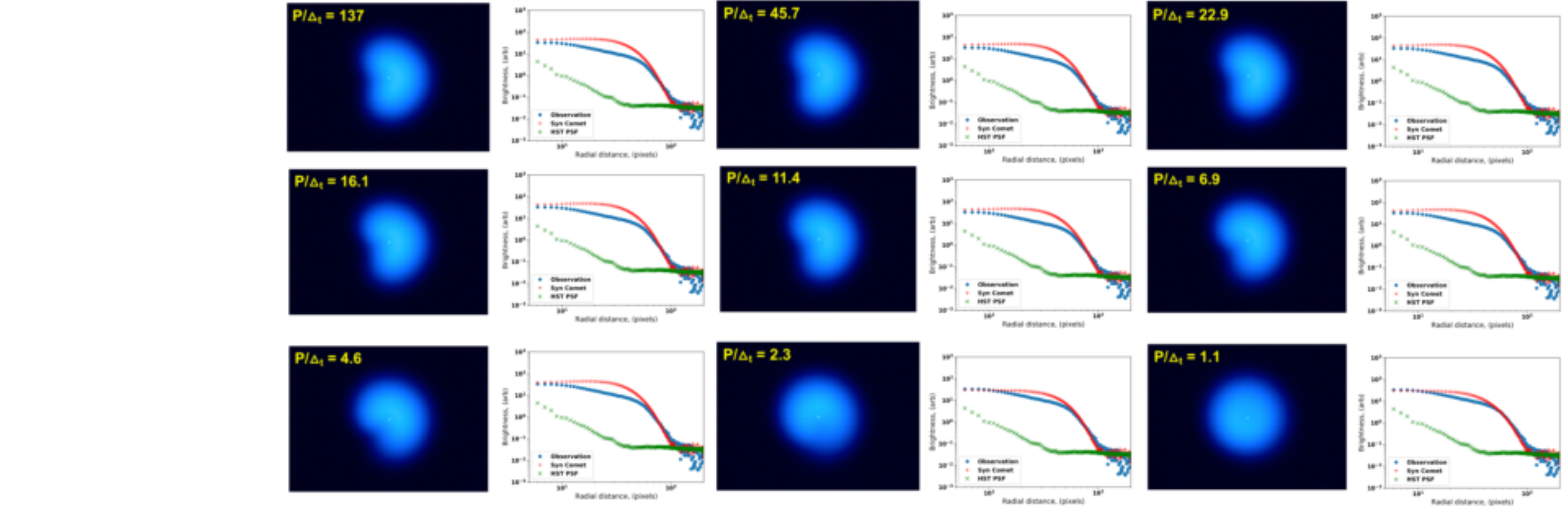}{0.9\textwidth}{}
          	}
\caption{Diagram is similar to Figure \ref{fig:spin_pole_1} and \ref{fig:spin_pole_2}, but for an assumed nucleus spin-pole direction directed along the nucleus-HST direction. For this panel, the top set of synthetic images and radial profiles are for Exp-1 and the bottom set are for Exp-2. When comparing Exp-1 and Exp-2 synthetic images through visual inspection and through radial surface brightness propfile comparison, we find a lower limit for P/$\Delta t_2$ is on the order of $\sim$ 7. \label{fig:spin_pole_4}}
\end{figure}

Results of the best-fit non-rotating nucleus model were used as the initial nucleus model input parameters for further modeling efforts to place lower-limit constraints on the nucleus' spin period. For our analysis we chose a grid of spin-pole orientations (right ascension and declination of the nucleus' angular momentum vector) for the nucleus which covered the 4$\pi$ steradian phase space. For each assumed spin-pole orientation, we proceeded to decrease the spin period from P = 60 days, while maintaining the modeled outburst duration, until the morphology of the synthetic images significantly deviated from the observation's morphology. This approach is similar to our earlier 2008 Kitt Peak 2.1-m analysis. Figures \ref{fig:spin_pole_1}, \ref{fig:spin_pole_2}, and \ref{fig:spin_pole_4} show examples of this modeling effort for three spin-pole orientations covering extremes in spin-pole orientation phase space.  	

For most spin-pole orientations there existed a lower limit to the ratio P/$\Delta t_2$ which could replicate the morphology detected in the observations. The lower limit varied depending on the assumed spin-pole orientation, but generally around a ratio P/$\Delta t_2$ $\leq$ $\sim$20-40 the coma morphology began to deviate significantly between observations and models. An assumed spin-pole directed along the nucleus-HST vector as depicted by Figure \ref{fig:spin_pole_4} provided one important exception to this trend. For this spin-pole orientation P/$\Delta t_2$ on the order of $\geq \sim$7 could replicate the observed morphology. 

To derive constraints on the nucleus' spin period an independent measure of the outburst duration must be known. Our earlier analysis (\citet{2017Icar..284..359S}) used the outburst coma's morphology to place an upper limit constraint on the outburst duration to $\sim$1.5 days. A similar analysis for the 1996 HST observations resulted in an upper limit to the outburst duration being on the order of a 5-10 hours depending on the specific combination of spin-pole orientation and spin period used. Lower limit constraints for the outburst duration could not be estimated directly from the outburst coma's morphology. Lower limit constraints can be derived by using the estimated amount of dust emitted during the outburst event (Section 3.3) if the outburst mechanism is known. Heretofore, no well understood mechanism has been determined which can explain SW1's outburst behavior. Derivations of outburst duration lower limit constraints through modeling the many proposed outburst mechanisms is beyond the scope of this current analysis. For our current work we use a range of plausible outburst durations to derive a range of spin period constraints assuming a typical P/$\Delta t_2$ = 20. Outburst durations on the order of 10 minutes, 1 hour, and 10 hours result in spin period lower limit constraints of 3.33 hours, 20 hours, and 8.33 days.

We emphasize that the modeling cases shown in Figures \ref{fig:spin_pole_2} and \ref{fig:spin_pole_4} were included to highlight spin-period constraints derived through a coma morphology analysis for extreme cases of spin-pole orientation parameter space. These spin-pole orientation extremes have a high probability of experiencing a large degree of seasonally-dependent activity behaviors. Long-baseline studies (observational epochs between 1996 to 2014) of SW1's quiescent and outburst activity behaviors do not show such extremes in seasonal activity behaviors; instead observations indicate a modestly varying outburst frequency close to $\sim$58 days/event along $\sim$75\% of its orbit (\cite{2016Icar..272..387M}). This observationally constrained outburst behavior for SW1 suggests that the two extreme spin-pole orientations shown are unlikely spin-pole orientations. 

\section{Conclusions and Discussion} \label{sec:conclusion}
The high-resolution HST WFPC2 observations of SW1 presented in this analysis showcase the level of inner comae morphology detectable with sufficient S/N for detailed 3-D Monte Carlo coma modeling efforts to characterize the 3-D outburst comae shapes and constrain nuclei spin states for distantly active Centaurs and comets. In particular, for SW1 we have characterized a multi-component coma present shortly after a UT March 1996 outburst. A 3-D Monte Carlo coma modeling effort of these observations resulted in constraints on the nucleus' source region of the outburst and the 3-D shape of the lofted dust grains. The model derived lower limits for the nucleus' spin period are in agreement with an earlier analysis by our group of a set of ground-based optical observations of a 2008 outburst of SW1 (\citet{2017Icar..284..359S}). Similarities in the inferred 3-D outburst coma shape between the UT March 1996 and September 2008 observations as derived through analysis of radial surface brightness profiles suggests evidence for common characteristics of SW1's frequent major (e.g. $\Delta m \geq   \sim$ 2-3) outbursts events. In particular: (1) the surface region of the outburst is small, on the order of 1-10\% of the nucleus' surface area. (2) The duration of time for increased dust lofting (outburst duration) from the surface is short when compared to the nucleus rotation period. We find that the ratio P/$\Delta t_2$ $\geq$ $\sim$ 20-40 depending on the modeling-assumed spin-pole orientation. (3) Constraints for an upper limit on the outburst duration as derived from the outburst coma's morphology are on the order of $\sim$10 $\pm$ 10 hours. (4) The 3-D outburst coma is typically a cone-shape jet originating near the sub-solar point on the nucleus' surface and directed along the Sun-nucleus vector. (5) The nucleus' rotation period is constrained to have a lower limit on the order of days for most plausible spin-pole orientations and strengthens the observational evidence that SW1 possesses one of the slowest rotation rates of known non-tidally locked solar system objects. While such a slow rotation state would be unique with regards to known rotation periods of JFCs (\cite{2017MNRAS.471.2974K}), similar slow rotation periods have been measured for a handful of small bodies (e.g. the Jupiter Trojan 11351 Leucus with an $\sim$19 day rotation period (\cite{2018AJ....155..245B})).

The lower limit for the amount of dust emitted during this outburst ((2.79 $\pm$ 0.05)$\times 10^8$ kg) is on the lower end of reported measurements for SW1: (2.6 $\pm$ 0.3)$\times 10^8$ kg (\citet{2013AJ....145..122H}) and (1.8 $\pm$ 0.07)$\times 10^9$ kg (\citet{2017Icar..284..359S}). Assuming a nucleus bulk density $\rho = 600$ kg/m$^3$ results in an effective spherical region excavated from SW1's nucleus with radius of $\sim$48 m.

\underline{To summarize:}
\begin{itemize}
\item We have characterized a complex multicomponent coma detected in two UT 1996 March observational epochs of HST broadband imaging of the Centaur SW1.
  
\item The UT 1996 March outburst dust coma was constrained to be fully contained within a 5$''$ aperture radius, equivalent to a projected nucleocentric skyplane distance of $\sim$19,000 km, and resulted in an apparent R-band magnitude of 12.86 $\pm$ 0.02. 

\item A sky-plane projected outflow velocity of 0.17 $\pm$ 0.05 km/s was measured for the outburst material contained in the 5$''$ aperture radius.

\item Photometry of the most recent outburst dust coma resulted in a lower limit constraint on the amount of dust emitted to be on the order of (2.79 $\pm$ 0.05)$\times 10^8$ kg.

\item The UT 1996 March outburst dust coma's 3-D shape was best fit by an asymmetric cone of material with apex directed along the nucleus-Sun vector and apex angle of $\sim$ (30 $\pm 5)^{\circ}$.

\item 3-D Monte Carlo coma modeling of the outburst dust coma's morphology contained within the 5$''$ radius aperture placed constraints on the nucleus' spin period on the order of days for all plausible spin-pole orientations.

\item Future investigations into understanding the mechanism(s) driving SW1's outbursts and the duration of increased dust lofting are necessary to better interpret the results of this 3-D Monte Carlo modeling effort and a similar modeling effort of UT September 2008 outburst observations of SW1. 

\end{itemize}

\acknowledgments
Based on observations made with the NASA/ESA Hubble Space Telescope, obtained from the Data Archive at the Space Telescope Science Institute, which is operated by the Association of Universities for Research in Astronomy, Inc, under NASA contract NAS 5-26555. These observations are associated with program \#5829.

Support for this work was provided by NASA through grant number AR-14294 from the Space Telescope Science Institute, which is operated by the Association of Universities for Research in Astronomy, Inc., under NASA contract NAS 5-26555. We also thank the NASA Earth \& Space Science Fellowship (NESSF) Grant \# NNX16AP41H and the Center for Lunar and Asteroid Surface Science (CLASS, NNA14AB05A) for additional support of this work.

We would like to thank the article's anonymous reviewer for their thorough review and helpful comments during the preparation of this article.

%





\bibliography{schambeau_refs}{}
\bibliographystyle{aasjournal}



\end{document}